\documentclass[12pt]{article}
\usepackage{geometry}                
\usepackage{graphicx}
\usepackage{amssymb}
\usepackage{epstopdf}
\usepackage{amsmath}
\usepackage{subcaption} 
\usepackage{booktabs} 
\usepackage{multirow} 

\usepackage{bm}
\usepackage{booktabs}
\usepackage[square,sort,comma,numbers]{natbib}
\newcommand{\nop}[1]{}
\usepackage{soul,color}
\usepackage{ragged2e}
\usepackage{booktabs,makecell,multirow,tabularx}
\usepackage{float}
\usepackage{hyperref}

\usepackage{enumitem}
\usepackage{makecell}

\usepackage{algorithm}
\usepackage{algpseudocode}

\usepackage{authblk}

\DeclareMathOperator*{\argmax}{argmax}

\title{Estimating the Joint Distribution of Two Binary Variables with Marginal Statistics}

\author{Longwen Shang}
\author{Min Tsao}
\author{Xuekui Zhang}
\affil{Department of Mathematics and Statistics, University of Victoria}

\date{May 6, 2025}

\begin{document}
\maketitle


\begin{abstract}
Clinical trial simulation (CTS) is critical in new drug development, providing insight into safety and efficacy while guiding trial design. Achieving realistic outcomes in CTS requires an accurately estimated joint distribution of the underlying variables. However, privacy concerns and data availability issues often restrict researchers to marginal summary-level data of each variable, making it challenging to estimate the joint distribution due to the lack of access to individual-level data or relational summaries between variables. We propose a novel approach based on the method of maximum likelihood that estimates the joint distribution of two binary variables using only marginal summary data. By leveraging numerical optimization and accommodating varying sample sizes across studies, our method preserves privacy while bypassing the need for granular or relational data. Through an extensive simulation study covering a diverse range of scenarios and an application to a real-world dataset, we demonstrate the accuracy, robustness, and practicality of our method. This method enhances the generation of realistic simulated data, thereby improving decision-making processes in drug development.
\end{abstract}

\section{Introduction}

Bringing a new drug to market is an exceptionally costly endeavor, with the average cost of developing and launching a new drug estimated at around \$2.6 billion \cite{nref1}. Additionally, the clinical trial process has a high failure rate, with approximately 90\% of drug development projects not advancing through all phases of clinical testing \cite{nref2, nref3}. To address these challenges, preliminary investigations using Clinical Trial Simulation (CTS) have become essential for mitigating risks and improving the chances of success. The FDA’s Complex Innovative Trial Designs (CID) program, introduced in response to U.S. legislation under the Prescription Drug User Fee Act, aims to reduce failure rates by incorporating novel and complex trial designs \cite{FDA2022}. Adaptive clinical trials, a key component of the CID program, are increasingly viewed as the future of clinical research due to their flexibility, efficiency, and potential to enhance patient outcomes. The FDA’s 2019 guidelines highlight the critical role of simulations in validating adaptive trial designs, particularly in evaluating statistical properties and ensuring the trials’ scientific rigor \cite{FDA2019}.

CTS has become a vital aspect of drug development, providing consistency in trial design, improving communication with stakeholders, and guiding adaptive decision-making processes \cite{ref10}. Mayer et al. (2019) emphasize that CTS is crucial in ensuring trial designs are well-structured and adaptable, which is essential for stakeholder engagement and informed decision-making \cite{Mayer2019}. Additionally, simulations are particularly valuable in reducing risks, especially in trials involving rare diseases with limited patient populations. Walker et al. (2020) illustrate how CTS can optimize trial power and resource allocation by refining eligibility criteria and follow-up schedules, thereby maximizing the efficiency of resource use \cite{Walker2020}. Westfall et al. (2008) discuss the importance of statistical templates in CTS, which accommodate complexities like patient dropout and allow for a thorough evaluation of different trial designs \cite{Westfall2008}. Similarly, Holford et al. (2010) highlight that CTS enables the integration of optimal design methodologies, enhancing the overall drug development process by providing a more robust evaluation framework \cite{Holford2010}.  From a financial perspective, simulations significantly reduce the number of necessary studies, increase the likelihood of trial success, and potentially shorten development timelines, which can lead to substantial cost savings for pharmaceutical companies \cite{ref9}. In an industry where speed and cost-efficiency are paramount, leveraging simulation studies offers a competitive advantage by streamlining the development process and ensuring more reliable trial outcomes.

Accurate simulation models and realistic parameter values are essential to ensure that clinical trial simulations are both effective and practical. However, obtaining patient-level data from populations that closely match the target demographic for a new drug is often unrealistic during the planning stage \cite{D4}. Using summary-level data from publications has become a popular approach because it provides a practical and accessible means to estimate key trial parameters without the need for detailed patient-level information. Data such as hazard ratios, event rates, and demographic characteristics are often readily available, making it an efficient resource for modelling and simulating new trials. These published summary statistics allow researchers to estimate critical factors like treatment effects, baseline event rates, and variability, which are crucial for trial design. The advantage of this approach is that it enables faster and more cost-effective simulations, especially when patient-level data is difficult to access or unavailable.

However, using summary-level data to simulate each variable in clinical trials has a critical limitation: it provides only marginal summaries of individual variables without capturing their joint distributions. Ignoring the joint distribution between demographic variables can lead to flawed simulations and inaccurate trial designs. For example, consider two binary demographic variables: gender (male/female) and smoking status (smoker/non-smoker). If we only use summary data, we might know the proportion of male and female patients, as well as the proportion of smokers and non-smokers, but not how these variables interact. In reality, there might be a strong correlation—perhaps most of the smokers in the study are male, and non-smokers are predominantly female. If this joint distribution is ignored, the simulation might assume an even distribution of smokers across genders, potentially leading to an inaccurate representation of the patient population. This could skew trial outcomes, particularly if smoking status has a significant impact on treatment efficacy or side effects. Ignoring such correlations may result in biased conclusions about the drug’s performance, particularly for specific subgroups, underscoring the importance of considering joint distributions when simulating clinical trials. 

Our research addresses a critical challenge in the field. This paper focuses on estimating the joint distribution of two binary variables using marginal summary data from multiple studies. Previous work has explored maximum likelihood estimation of the parameters of a bivariate binomial distribution. Hamdan et al. (1971) \cite{ref2} studied estimation of the joint distribution of two binary variables based on individual-level data, but their result is not applicable when only summary-level data is available. A follow-up study \cite{ref1} extended this work to maximum likelihood estimation using marginal summary data from multiple studies, but it relied on the assumption of equal sample sizes across studies, which is unrealistic in real-world applications where sample sizes often vary. Building on previous research, we propose a novel method that generalizes the maximum likelihood approach for estimating the bivariate binomial distribution. Our method accommodates marginal summary data from studies with varying sample sizes. Additionally, we quantify uncertainty in our estimates using Fisher Information and provide both point and interval estimates. This work advances the theoretical framework for privacy-preserving statistical inference and offers practical tools handle summary-level data sources.

The remainder of this paper is organized as follows. Section 2 introduces the theoretical derivation of the likelihood function, the observed Fisher Information, and the optimization framework. Section 3 evaluates the proposed method using simulation studies across diverse data scenarios. Section 4 applies the method to real-world data from 12 cancer-related datasets, demonstrating its practical utility. Section 5 concludes with a discussion on future research.


\section{Method}
\subsection{Problem Setup and Notation}

Estimating the joint distribution of two binary variables from marginal summaries of datasets is a challenging problem. It is unsolvable when only a single dataset is available as the marginal summaries of a single dataset contains no information about the correlation of the two variables. However, when there are multiple datasets drawn independently from the same population—and thus share a common underlying joint probability mass function (PMF)—it becomes feasible to estimate the joint distribution. Below, we introduce the necessary notation and formally define this problem.

\textbf{Joint distribution}
Let \( I_x \sim \text{Bernoulli}(p_1) \) and \( I_y \sim \text{Bernoulli}(p_2) \) represent two binary random variables.  The joint distribution of \( I_x \) and \( I_y \) can be parameterized by three parameters: \( (p_1, p_2, p_{11}) \), where \( p_{11} = P(I_x = 1, I_y = 1) \). Table \ref{table1} gives the joint PMF of \( I_x \) and \( I_y \), along with their marginal PMFs in the last row and last column.

\begin{table}[ht]
\centering
\caption{Joint and Marginal Distributions of Two Binary Variables}
\label{table1}
\begin{tabular}{cccc}
  \toprule
             & $I_y = 1$      & $I_y = 0$              & Totals \\
  \midrule
  $I_x = 1$  & $p_{11}$       & $p_{1}-p_{11}$         & $p_{1}$ \\
  $I_x = 0$  & $p_{2}-p_{11}$ & $1-p_{1}-p_{2}+p_{11}$ & $1-p_{1}$ \\
  Totals     & $p_{2}$        & $1-p_{2}$              & $1$ \\
  \bottomrule
\end{tabular}
\label{probtable}
\end{table}

\textbf{Observed Data from Multiple Datasets}

Consider \( k \) independent datasets (samples), all drawn from the same population with a joint PMF given in Table~\ref{table1}. For the \( i \)-th dataset (\( i = 1, \ldots, k \)), let \( n_i \) denote the sample size, \( x_i \) denote the number of observations where \( I_x = 1 \), \( y_i \) denote the number of observations where \( I_y = 1 \), and \( z_i \) denote the number of observations where both \( I_x = 1 \) and \( I_y = 1 \) (so $0\leq z_i\leq \min\{x_i, y_i\}$). These quantities form the \( i \)-th dataset shown in Table~\ref{BBDtable} below.

\begin{table}[ht]
\centering
\caption{Bivariate Binomial Contingency Table for the \( i \)-th Dataset}
\label{BBDtable}
\begin{tabular}{cccc}
  \toprule
             & \( I_y = 1 \)  & \( I_y = 0 \)  & Totals \\
  \midrule
  \( I_x = 1 \)  & \( z_i \)        & \( x_i - z_i \)      & \( x_i \) \\
  \( I_x = 0 \)  & \( y_i - z_i \)    & \( n_i - x_i - y_i + z_i \)  & \( n_i - x_i \) \\
  Totals     & \( y_i \)        & \( n_i - y_i \)      & \( n_i \) \\
  \bottomrule
\end{tabular}
\end{table}
The marginal summaries or summary-level data refer to marginal totals $(x_i, y_i, n_i)$ in Table~\ref{BBDtable}. In the present discussion, only such summaries are observed.

\textbf{Problem Formulation}
Our objective is to estimate the joint PMF of \( I_x \) and \( I_y \), parameterized by \( (p_1, p_2, p_{11}) \), using the observed data \( \{(x_i, y_i, n_i); i = 1, \ldots, k\} \). Since the \( z_i \)'s are latent variables whose values cannot be observed, the estimation process involves maximizing the likelihood of the model after marginalizing over \( z_i \). This approach makes use of all $k$ datasets to estimate \( (p_1, p_2, p_{11}) \).

\subsection{Likelihood}

The joint distribution of two binary variables given in Table~\ref{table1} may be viewed as a multinomial distribution having four possible outcomes with the following probabilities:
\begin{align*}
    P(I_x = 1, I_y = 1) &= p_{11}, \\
    P(I_x = 1, I_y = 0) &= p_1 - p_{11}, \\
    P(I_x = 0, I_y = 1) &= p_2 - p_{11}, \\
    P(I_x = 0, I_y = 0) &= 1 - p_1 - p_2 + p_{11}.
\end{align*}

By Table~\ref{BBDtable}, the observed counts corresponding to these outcomes in the \( i \)-th dataset are \( z_i \), \( x_i - z_i \), \( y_i - z_i \), and \( n_i - x_i - y_i + z_i \), respectively. Thus, the joint PMF of two binary variables is equivalent to the PMF of the multinomial distribution. 

Consistent with that in \cite{ref1}, the `complete-data' likelihood of the \( i \)-th dataset is:

\begin{equation}
    T(x_i, y_i, z_i, n_i; \bm{p}) = \frac{n_i! \cdot p_{11}^{z_i} (p_1 - p_{11})^{x_i - z_i} (p_2 - p_{11})^{y_i - z_i} (1 - p_1 - p_2 + p_{11})^{n_i - x_i - y_i + z_i}}{z_i! (x_i - z_i)! (y_i - z_i)! (n_i - x_i - y_i + z_i)!},
\end{equation}
where \( \bm{p} = (p_1, p_2, p_{11}) \). The `observed-data' likelihood is derived by marginalizing over latent variable \( z_i \)'s as follows:
\begin{equation}
    b(x_i, y_i, n_i; \bm{p}) = \sum_{z_i=0}^{\min(x_i, y_i)} T(x_i, y_i, z_i, n_i; \bm{p}).
    \label{pmf}
\end{equation}
So, the log-likelihood of $\bm{p}$ based on the observed marginal summary data from all \( k \) datasets is:

\begin{align} \label{eq:loglik}
    l(\bm{p}) = \sum_{i=1}^k \log b(x_i, y_i, n_i; \bm{p}).
\end{align}

\subsection{Maximum Likelihood Estimation (MLE) }
The MLE of $(p_1, p_2, p_{11})$ is the solution of the following system of equations derived by setting the partial derivatives of the log-likelihood function to zero:
\begin{subequations}\label{eq}
\begin{align}
\frac{\partial l(\bm{p})}{\partial p_1} &= \frac{1 - p_2}{p_1 - p_{11}} \sum_i x_i + \sum_i y_i - \sum_i n_i - \frac{1 - p_2}{p_1 - p_{11}} S = 0 \label{lf1} \\
\frac{\partial l(\bm{p})}{\partial p_2} &= \frac{1 - p_1}{p_2 - p_{11}} \sum_i y_i + \sum_i x_i - \sum_i n_i - \frac{1 - p_1}{p_2 - p_{11}} S = 0 \label{lf2} \\
\frac{\partial l(\bm{p})}{\partial p_{11}} &= \sum_i x_i + \sum_i y_i - \sum_i n_i + \frac{1 - p_1 - p_2}{p_{11}} S = 0 \label{lf3}
\end{align}
\end{subequations}

Here, $S$ represents the expectation of the latent variable $z_i$ given the observed data $(x_i, y_i, n_i)$ and the current parameter estimates, and is given by 
\begin{equation}
    S(x_i, y_i, n_i; \bm{p}) = \sum_{i=1}^k \sum_{z_i} \frac{z_i \cdot T(x_i, y_i, z_i, n_i;\bm{p})}{b(x_i, y_i, n_i;\bm{p})}.
\end{equation}

We now briefly explain how to obtain the solution for this system of equations. Detailed steps may be found in Appendix~\ref{appendix:MLE}.

Closed-form solutions (i.e., MLEs) for $p_1$ and $p_2$ can be derived from the system of equations~(\ref{eq}), and they are
\begin{align} \label{eq:MLEp1p2}
\hat{p}_1 = \frac{\sum_i x_i}{\sum_i n_i} \quad \text{and} \quad
\hat{p}_2 = \frac{\sum_i y_i}{\sum_i n_i}.
\end{align}

However, there is no closed-form solution for the MLE of $p_{11}$. We thus substitute $\hat{p}_1$ and $\hat{p}_2$ from equation \eqref{eq:MLEp1p2} into the log-likelihood \eqref{eq:loglik}, so the MLE of $p_{11}$ is now the solution of the following univariate optimization problem: 
\begin{align}\label{eq:p11optim}
    \hat{p}_{11} = \argmax \sum_{i=1}^k \log b(x_i,y_i,n_i;\frac{\sum_i x_i}{\sum_i n_i}, \frac{\sum_i y_i}{\sum_i n_i}, p_{11}),      
\end{align}
where $0\leq p_{11}\leq \min\{\hat{p}_1,\hat{p}_2\}$. This boundary-constrained optimization problem can be solved using the L-BFGS-B algorithm (Limited-memory Broyden-Fletcher-Goldfarb-Shanno with Box constraints) based R package \texttt{optim}. This quasi-Newton method iteratively approaches the optimal solution using gradient information, while efficiently handling the parameter boundary constraints by storing only a limited amount of past update information. The initial value for L-BFGS-B algirthm is chosen through a grid search over the interval $[0,1]$.

\subsection{Standard Errors and Interval Estimation I: The Normal Approximation Approach}

A key advantage of the maximum likelihood approach to parameter estimation is that the MLE is in general consistent and asymptotically normally distributed. These properties make it possible to conduct hypothesis tests and construct confidence intervals (CI) for the unknown parameter, provided the standard error (SE) of the MLE can be effectively estimated. For the present problem, the 100\((1-\alpha)\)\% normal approximation based CIs for $p_1, p_2$ and $p_{11}$ are

\begin{align}
    CI_1({p}_{1}) &= \left( \hat{p}_{1} - z_{1 - \frac{\alpha}{2}} \cdot \hat{\textrm{SE}}(\hat{p}_{1}), \;\; \hat{p}_{1} + z_{1 - \frac{\alpha}{2}} \cdot \hat{\textrm{SE}}(\hat{p}_{1}) \right), \\
    CI_1({p}_{2}) &= \left( \hat{p}_{2} - z_{1 - \frac{\alpha}{2}} \cdot \hat{\textrm{SE}}(\hat{p}_{2}), \;\; \hat{p}_{2} + z_{1 - \frac{\alpha}{2}} \cdot \hat{\textrm{SE}}(\hat{p}_{2}) \right), \\
    CI_1({p}_{11}) &= \left( \hat{p}_{11} - z_{1 - \frac{\alpha}{2}} \cdot \hat{\textrm{SE}}(\hat{p}_{11}), \;\; \hat{p}_{11} + z_{1 - \frac{\alpha}{2}} \cdot \hat{\textrm{SE}}(\hat{p}_{11}) \right),
    \label{eq:CI1}
\end{align}

respectively, where $z_{1 - \frac{\alpha}{2}}$ is the $(1 - \frac{\alpha}{2})$th quantile of a standard normal distribution and $\hat{\textrm{SE}}(\hat{p}_{1})$, for example, denotes the estimated SE of $\hat{p}_1$.

To estimate the SEs of \(\hat{p}_{1}\) and \(\hat{p}_{2}\), noting that they are each the sample proportion of a combined sample of size $\sum_i {n_i}$ formed by the $k$ independent datasets from the same population, their estimated SEs are, respectively,
\[
\hat{\textrm{SE}}(\hat{p}_{1}) = \sqrt{\frac{\hat{p}_{1}(1-\hat{p}_{1})}{\sum_i n_i}} \quad \textrm{and} \;\; \hat{\textrm{SE}}(\hat{p}_{2}) = \sqrt{\frac{\hat{p}_{2}(1-\hat{p}_{2})}{\sum_i n_i}}.
\]

Since $\hat{p}_{11}$ is not a sample proportion and it does not have an analytic expression, we estimate its SE via the observed Fisher information as follows:

\begin{align} \label{e:p11SE}
\hat{\textrm{SE}}(\hat{p}_{11}) = 
\sqrt{\mathcal{I}(\hat{p}_{11}; x_i, y_i, n_i)^{-1}}
=
\sqrt{
\left( -
\frac{\partial^2 l(\hat{\bm{p}})}{\partial \hat{p}_{11}^2}
\right)^{-1}
}.
\end{align}

Here, \(\mathcal{I}(\hat{p}_{11}; x_i, y_i, n_i)\) represents the observed Fisher information for $\hat{p}_{11}$. The second derivative of the log-likelihood function in (\ref{e:p11SE}) has a closed-form expression shown in (\ref{fisherinfo}), so the estimated SE in (\ref{e:p11SE}) can be computed via (\ref{fisherinfo}). The SEs of $\hat{p}_1$ and $\hat{p}_2$ may also be estimated via their Fisher information but the estimates given above are more accurate.

\begin{equation}
\frac{\partial^2 l({\bm{p}})}{\partial {p}_{11}^2} = 
\sum_{i=1}^k \frac
{b(x_i, y_i, n_i) \cdot \sum_{z_i} \left\{ T(\mathbf{v}_i) \left[ \Delta^2(\mathbf{v}_i) - \Lambda(\mathbf{v}_i) \right] \right\} - \left[\sum_{z_i} T(\mathbf{v}_i) \cdot \Delta(\mathbf{v}_i)\right]^2}
{[b(x_i, y_i, n_i)]^2},
\label{fisherinfo}
\end{equation}

where $\mathbf{v}_i = (z_i, x_i, y_i, n_i)$,

\[
\Delta(\mathbf{v}_i) = \frac{z_i}{p_{11}} - \frac{x_i - z_i}{p_1 - p_{11}} - \frac{y_i - z_i}{p_2 - p_{11}} + \frac{n_i - x_i - y_i + z_i}{1 - p_1 - p_2 + p_{11}},
\]

\[
\Lambda(\mathbf{v}_i) = \frac{z_i}{p_{11}^2} + \frac{x_i - z_i}{(p_1 - p_{11})^2} + \frac{y_i - z_i}{(p_2 - p_{11})^2} + \frac{n_i - x_i - y_i + z_i}{(1 - p_1 - p_2 + p_{11})^2}.
\]

\subsection{Interval Estimation II: The Likelihood Ratio Approach}

The normal approximation may not be accurate when the number of datasets, \( k \), is small. Additionally, normal approximation based CIs may violate the bounds on the parameters, e.g., the normal approximation based CI for $p_1$ may contain values outside of the interval \([0, 1]\). To address these issues, we propose an alternative approach to construct CIs for the estimated parameters using the likelihood ratio test.

The likelihood ratio test statistic for \(H_0: p=p_{11} \) is
\[
\lambda_{11}(p) = -2 \left[ l(\hat{p}_1, \hat{p}_2, p) - l(\hat{p}_1, \hat{p}_2, \hat{p}_{11}) \right],
\]
and \( \lambda_{11}(p) \) has an asymptotic Chi-square distribution with one degree of freedom under $H_0$. It follows that an asymptotic \(100(1 - \alpha)\%\) CI for \( {p}_{11} \) is:
\begin{equation}
CI_2({p}_{11}) = \left\{ p : \lambda_{11}(p) \leq \chi^2_{1, 1-\alpha} \right\}.
\label{eq:CI2}
\end{equation}

where $\chi^2_{1, 1-\alpha}$ is the $(1-\alpha)$th quantile of the Chi-sqaure distribution with one degree of freedom.
The lower and upper endpoints of this CI can be obtained by solving the equation \( \lambda_{11}(p) = \chi^2_{1, 1-\alpha} \) numerically for its two roots which are the endpoints (see Figure \ref{ci_1} for an example). 

Similarly, the likelihood ratio test based CIs for $p_1$ and $p_2$ are:
\begin{align}
    CI_2({p}_{1}) \!=\! \left\{ p \!: -2 \left[ l(p, \hat{p}_2, \hat{p}_{11}) \!-\! l(\hat{p}_1, \hat{p}_2, \hat{p}_{11}) \right] \leq \chi^2_{1, 1-\alpha} \right\},\\
    CI_2({p}_{2}) \!=\! \left\{ p \!: -2 \left[ l(\hat{p}_1, p, \hat{p}_{11}) \!-\! l(\hat{p}_1, \hat{p}_2, \hat{p}_{11}) \right] \leq \chi^2_{1, 1-\alpha} \right\}.
\end{align}


\section{Simulation Studies}

In this section, we evaluate the effectiveness of our proposed method through a comprehensive simulation study. We focus on point and interval estimates for $p_{11}$ which are new. The point estimators for $p_1$ and $p_2$ coincide with sample proportions whose properties are well understood.

\subsection{Simulation design}
We consider twenty scenarios varying along three dimensions: correlation strength between variables, sample sizes $n_i$, and the total number of studies (datasets) $k$.

The Phi coefficient ($\Phi$) is a measure of dependency between two binary variables commonly used for 2x2 contingency tables. It is defined as follows:
$$\Phi = \frac{p_{11} - p_1 p_2}{\sqrt{p_1 (1 - p_1) \cdot p_2 (1 - p_2)}}.$$
Similar to the Pearson Correlation Coefficient, $\Phi$ has a range of $[-1,1]$ with $\Phi=-1$ representing perfect negative correlation and  $\Phi=1$ representing perfect positive correlation. When $\Phi=0$, the two binary variables are not only uncorrelated but also independent. The twenty scenarios involve two levels of correlation strength: (i) $\Phi=0.171$ representing weak correlation and (ii) $\Phi=0.796$ representing strong correlation. These two levels are obtained through two convenient sets of parameter values shown in Table \ref{simulation_scheme}. To explore the impact of the number of studies $k$, we included five $k$ levels: 10, 20, 30, 40, and 50. For sample sizes of the datasets, we considered two cases with $n_i$ values randomly selected from (a) $\{800, 801, \dots, 1000\}$ representing large sample-size studies, and (b) $\{100, 101, \dots, 200\}$ representing small sample-size studies.

\begin{table}[ht]
\centering
\caption{Correlation value $\Phi$ and number of studies $k$ for the 20 scenarios in the simulation study. The two $\Phi$ values correspond to the two convenient parameter settings in the second column.}
\label{simulation_scheme}
\begin{small}
\begin{sc}
\begin{tabular}{ccc}
\toprule
{Correlation} & {Parameters} & {Sample Sizes} \\
\midrule
\multirow{3}{*}{Weak (0.171)} & $p_{11} = 0.25$ & \multirow{6}{*}{$k = 10, 20, 30, 40, 50$} \\
                              & $p_1 = 0.35$    &                                 \\
                              & $p_2 = 0.60$    &                                 \\
\cmidrule(lr){1-2}
\multirow{3}{*}{Strong (0.796)} & $p_{11} = 0.75$ &                                 \\
                                & $p_1 = 0.80$    &                                 \\
                                & $p_2 = 0.77$    &                                 \\
\bottomrule
\end{tabular}
\end{sc}
\end{small}
\end{table}

\subsection{Data Generation Algorithm}

The summary-level datasets $(x_i, y_i, n_i)$ used in our simulation study are generated by Algorithm~\ref{alg:simulation}. To generate such a dataset for a scenario, we simply set the input parameter values to those of that scenario.

\begin{algorithm}[tb]
   \caption{Simulating Summary-level Sample Data}
   \label{alg:simulation}
\begin{algorithmic}[1]
   \State \textbf{Input:} $k$, $n_{\min}$, $n_{\max}$, $p_1$, $p_2$, $p_{11}$
   \For{$i = 1$ to $k$}
       \State Randomly select $n_i$ from $\{n_{\min}, n_{\min}+1, \dots, n_{\max}\}.$
       \State Simulate $(z_i, x_i - z_i, y_i - z_i, n_i - x_i - y_i + z_i)$
       \State \hspace{1em} $\sim \text{Multinomial}(n_i; p_{11}, p_1 - p_{11}, p_2 - p_{11}, 1 - p_1 - p_2 + p_{11})$
       \State \hspace{1em} \textit{as outlined in Table~\ref{table1} and Table~\ref{BBDtable}.}
       \State Calculate $x_i = z_i + (x_i - z_i)$ and $y_i = z_i + (y_i - z_i).$
   \EndFor
   \State \textbf{Output:} $(n_i, x_i, y_i)$ for $i = 1, \dots, k$
\end{algorithmic}
\end{algorithm}

\subsection{Performance of point and interval estimators}

For each scenario, we performed 1000 simulation runs, generating 1000 independent datasets using Algorithm~\ref{alg:simulation}. We computed 1000 independent estimates of $p_{11}$ and their estimated SEs with these datasets using the methodology outlined in Section 2 and R package \texttt{optim}. 

\begin{figure}[ht]
    \centering
    \includegraphics[width=\linewidth]{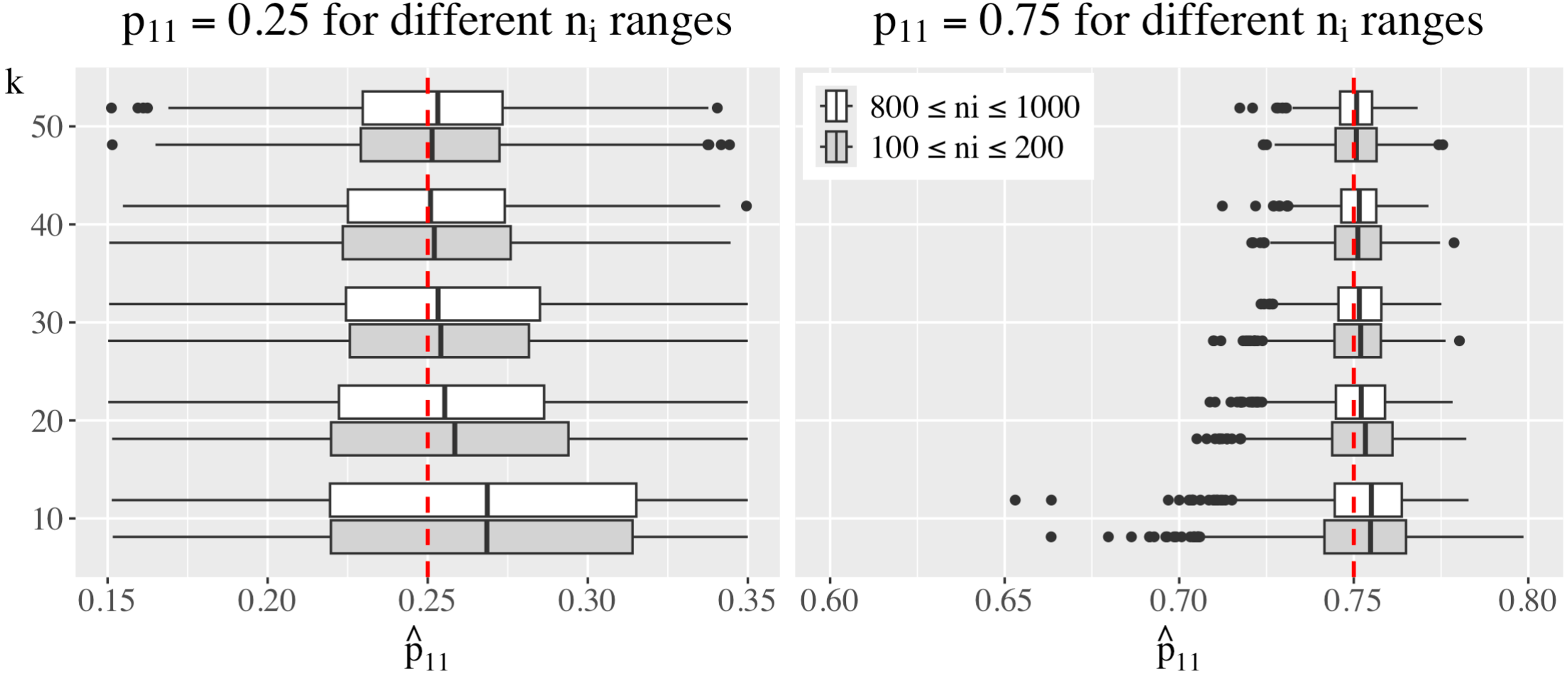}
    \caption{Boxplots of estimated $p_{11}$ values. Each boxplot is drawn with 1000 estimated values for a scenario. The red lines indicate the true $p_{11}$ values of 0.25 (left panel) and 0.75 (right panel).}
    \label{estimation_results}
\end{figure}

The 1000 estimates of $p_{11}$ for each scenario are summarized through a boxplot in Figure \ref{estimation_results}. The figure shows that the estimation precision improves in terms of both median error (the distance between the mid-line of a boxplot and the red vertial line) and variability (in terms of the spread captured by the boxplot) as $k$ increases. The impact of $n_i$ range is similar but less pronounced when compared to that of $k$.

To assess the adequacy of the normal approximation to the finite sample distribution of $\hat{p}_{11}$, we generated the histogram and QQ plot for the 1000 $\hat{p}_{11}$ values for each scenario. These plots (see Appendix \ref{appendix:normal}) show that, except for cases where $k=10$, the normal approximation is adequate. For cases where $k=10$, the histograms are quite left-skewed, and the QQ plots also show clear deviations from a line, indicating that the normal approximation is not appropriate.

\begin{figure}[ht]
    \centering
    \includegraphics[width=\linewidth]{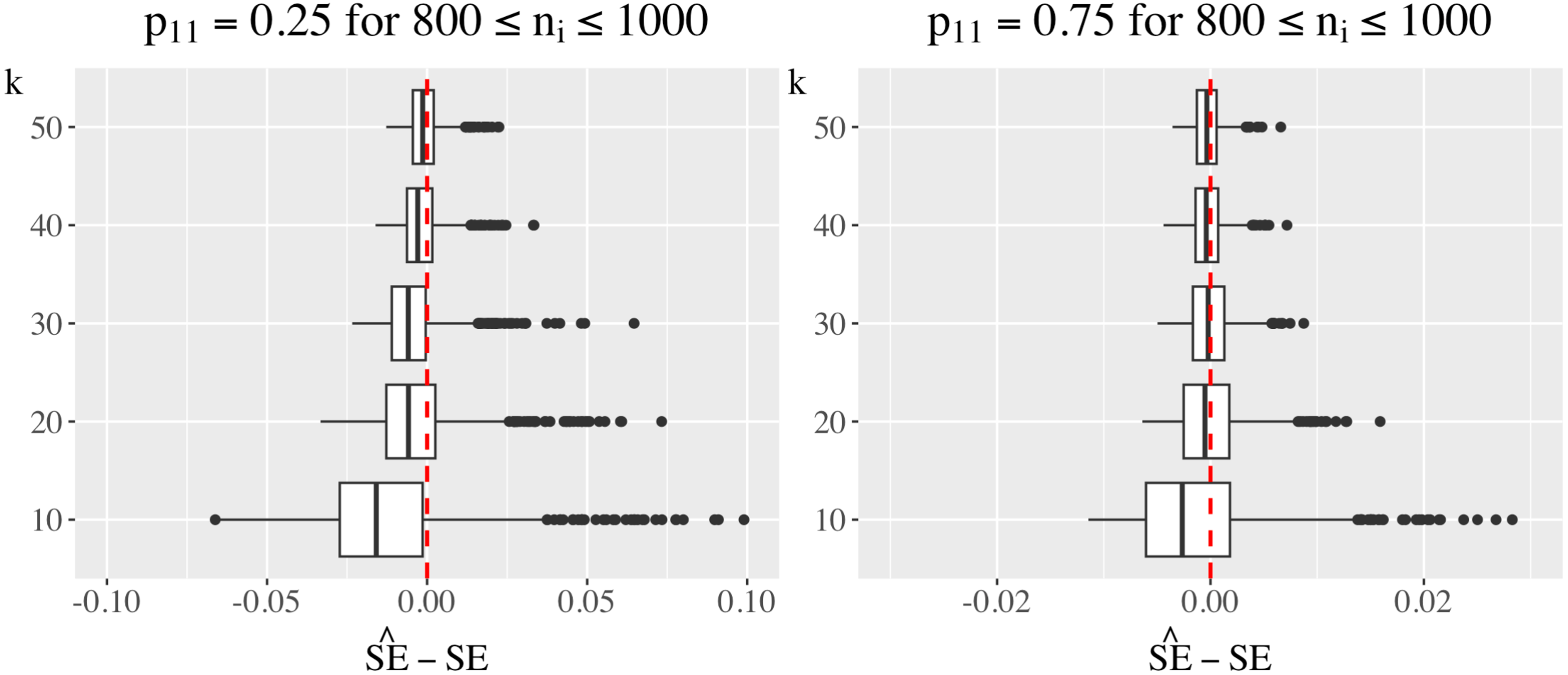}
    \caption{Boxplots of bias ($\hat{SE}$ $-$ true SE) of the estimated standard error for 10 scenarios. The red line represents $y=0$.}
    \label{ab_diff_se_a}
\end{figure}

For each scenario, we also examined the estimated standard error $(\hat{SE})$ of $\hat{p}_{11}$, which was calculated using the observed Fisher information. Figure \ref{ab_diff_se_a} shows the boxplot of the bias of the 1000 $\hat{SE}$ values for cases where $800 \leq n_i \leq 1000$. In order to compute these bias values, we needed the true SE of $\hat{p}_{11}$ which is unknown. We thus used the sample standard deviation of the large sample of 1000 simulated $\hat{p}_{11}$ values in its place. The figure shows that the bias is small relative to $p_{11}$, especially for larger $k$ values, confirming that the Fisher information based estimator $\hat{SE}$ is reliable. Results for cases where $100 \leq n_i \leq 200$ are similar, and they are provided in Appendix~\ref{appendix:normal}.

Finally, we computed confidence intervals for ${p}_{11}$ using both the normal approximation approach (\ref{eq:CI1}) and the likelihood ratio approach (\ref{eq:CI2}). Based on the joint probability Table~\ref{table1}, we identify the following constraints on parameters:

$$
\left\{
\begin{aligned}
& 1 \geq p_1 \geq p_{11} \geq 0 \\
& 1 \geq p_2 \geq p_{11} \geq 0 \\
& 1 \geq 1 - p_1 - p_2 + p_{11} \geq 0
\end{aligned}
\right.
$$
Confidence intervals computed using (\ref{eq:CI1}) and (\ref{eq:CI2}) are updated to ensure that they do not violate the above constraints. Such updated confidence intervals are compared below. 

\begin{table}[ht]
    \centering
    \caption{Coverage rates of 95\% confidence intervals for different values of $p_{11}$ and sample size ranges.}
    \label{tab:coverage_rates}
    \begin{tabular}{c cc cc cc cc}
        \toprule
            & \multicolumn{4}{c}{$p_{11} = 0.25$} & \multicolumn{4}{c}{$p_{11} = 0.75$} \\ 
        \cmidrule(lr){2-5} \cmidrule(lr){6-9}
        $k$ & \multicolumn{2}{c}{$100 \leq n_i \leq 200$} & \multicolumn{2}{c}{$800 \leq n_i \leq 1000$} 
            & \multicolumn{2}{c}{$100 \leq n_i \leq 200$} & \multicolumn{2}{c}{$800 \leq n_i \leq 1000$} \\
        \cmidrule(lr){2-3} \cmidrule(lr){4-5} \cmidrule(lr){6-7} \cmidrule(lr){8-9}
            & \(CI_1\) & \(CI_2\) & \(CI_1\) & \(CI_2\) & \(CI_1\) & \(CI_2\) & \(CI_1\) & \(CI_2\) \\ 
        \midrule
        10  & 0.863 & 0.916 & 0.859 & 0.930 & 0.852 & 0.862 & 0.861 & 0.933 \\
        20  & 0.856 & 0.933 & 0.882 & 0.944 & 0.841 & 0.882 & 0.900 & 0.940 \\
        30  & 0.906 & 0.934 & 0.890 & 0.931 & 0.848 & 0.870 & 0.880 & 0.935 \\
        40  & 0.908 & 0.940 & 0.914 & 0.948 & 0.861 & 0.900 & 0.884 & 0.936 \\
        50  & 0.935 & 0.949 & 0.927 & 0.951 & 0.868 & 0.883 & 0.921 & 0.946 \\
        \bottomrule
    \end{tabular}
\end{table}

\begin{table}[ht]
    \centering
    \caption{Average widths of 95\% confidence intervals for different values of $p_{11}$ and sample size ranges.}
    \label{tab:average_widths}
    \begin{tabular}{c cc cc cc cc}
        \toprule
            & \multicolumn{4}{c}{$p_{11} = 0.25$} & \multicolumn{4}{c}{$p_{11} = 0.75$} \\ 
        \cmidrule(lr){2-5} \cmidrule(lr){6-9}
        $k$ & \multicolumn{2}{c}{$100 \leq n_i \leq 200$} & \multicolumn{2}{c}{$800 \leq n_i \leq 1000$} 
            & \multicolumn{2}{c}{$100 \leq n_i \leq 200$} & \multicolumn{2}{c}{$800 \leq n_i \leq 1000$} \\
        \cmidrule(lr){2-3} \cmidrule(lr){4-5} \cmidrule(lr){6-7} \cmidrule(lr){8-9}
            & \(CI_1\) & \(CI_2\) & \(CI_1\) & \(CI_2\) & \(CI_1\) & \(CI_2\) & \(CI_1\) & \(CI_2\) \\ 
        \midrule
        10  & 0.226 & 0.213 & 0.224 & 0.212 & 0.053 & 0.070 & 0.051 & 0.070 \\
        20  & 0.188 & 0.173 & 0.189 & 0.175 & 0.040 & 0.044 & 0.041 & 0.046 \\
        30  & 0.160 & 0.150 & 0.161 & 0.149 & 0.034 & 0.036 & 0.035 & 0.036 \\
        40  & 0.142 & 0.133 & 0.142 & 0.133 & 0.030 & 0.031 & 0.030 & 0.031 \\
        50  & 0.126 & 0.119 & 0.127 & 0.120 & 0.027 & 0.028 & 0.027 & 0.028 \\
        \bottomrule
    \end{tabular}
\end{table}

Table~\ref{tab:coverage_rates} and Table~\ref{tab:average_widths} present the coverage rates and average widths of 95\% normal-based confidence intervals (\( CI_1 \)) and likelihood ratio-based confidence intervals (\( CI_2 \)) under different settings over 1000 runs. Both methods exhibit an under-coverage issue; the average coverage rate of \( CI_2 \) is 92.32\%, whereas that of \( CI_1 \) is only 88.28\%. Generally, the coverage rate improves when \( k \) increases. There are, however, exceptions where a higher \( k \) does not always guarantee a higher observed coverage rate. In terms of average width, \( CI_2 \) produces narrower intervals compared to \( CI_1 \) when the correlation is low ($p_{11}=0.25$). When the correlation is high ($p_{11}=0.75$), the widths of both methods are comparable, but \( CI_2 \) consistently achieves a higher coverage rate than \( CI_1 \). This indicates that \( CI_2 \) is more efficient. The inferior performance of \( CI_1 \) may be due to its dependence on the estimated SE. As shown in Figure~\ref{ab_diff_se_a} that the observed Fisher information approach tends to underestimate the SE, leading to narrower confidence intervals and consequently lower coverage rates. With larger $k$ and \( n_i \), the under-coverage issue is mitigated, and the likelihood ratio-based confidence intervals approach the nominal level of 95\%. 

\begin{figure}[ht]
    \centering
    \includegraphics[width=\linewidth]{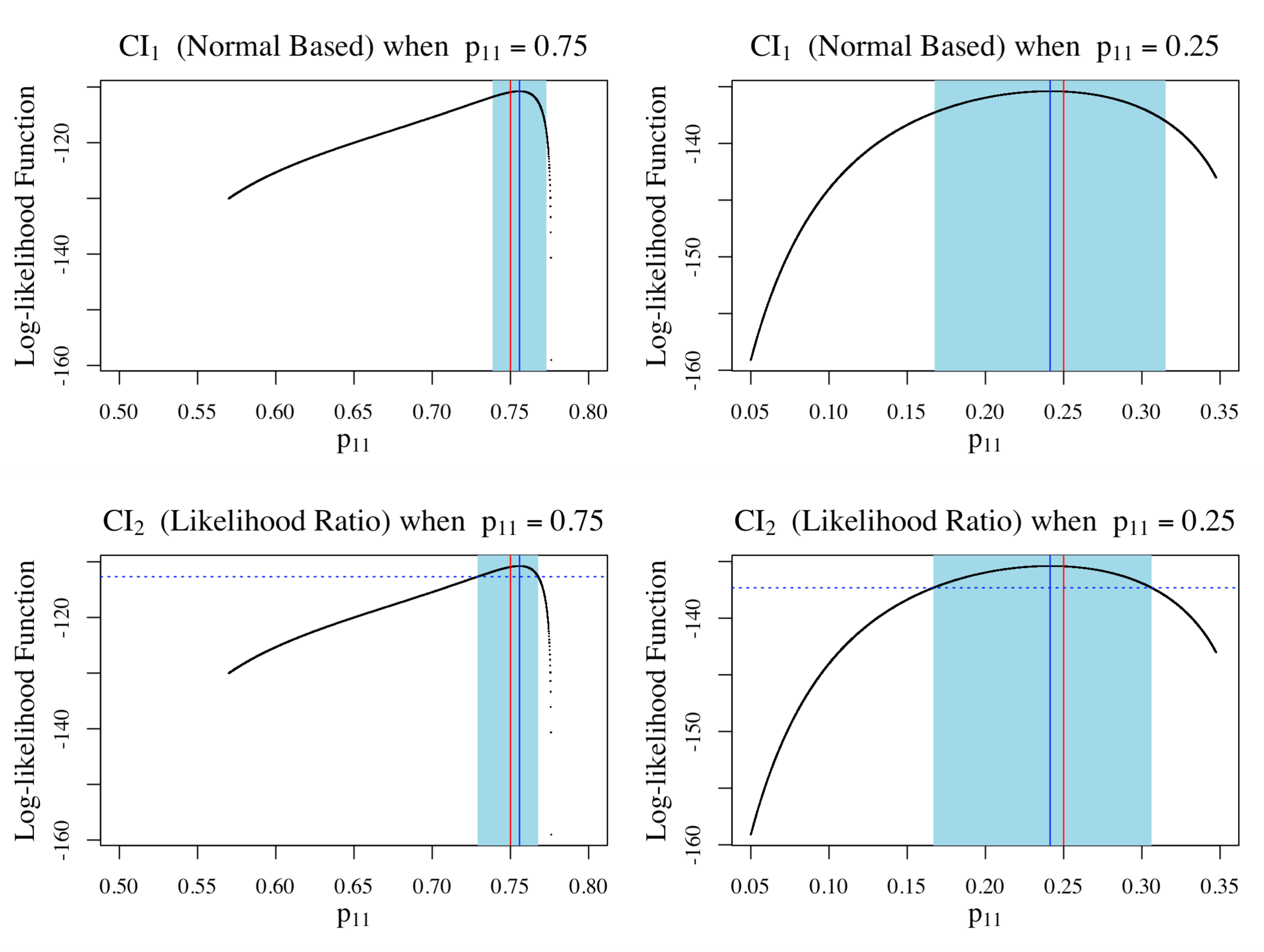} 
    \caption{Normal approximation vs likelihood ratio based 95\% confidence intervals. True values are $p_{11} = 0.75$ (left) and $p_{11} = 0.25$ (right). Red line indicates the true value, blue line represents $\hat{p}_{11}$, and the base of a blue-band is the confidence interval.}
    \label{ci_1}
\end{figure}

As an example, Figure~\ref{ci_1} illustrates the 95\% confidence intervals obtained from both methods when the true values of \( p_{11} \) are 0.25 and 0.75, respectively, with a total sample size of \( k = 20 \) and \( 100 \leq n_i \leq 200 \).

\subsection{Effect of $k$ and $n_i$ on Estimation Results}

During simulation study, there were occasions when the Observed Fisher Information approach gave negative estimates of the variance of $\hat{p}_{11}$, so the estimated SEs were unavailable. For small sample scenarios of $k=10$, such events occur less than 1\% of the time. For larger $k$ scenarios, they are very rare. To handle such events, our computing code sets the estimated SE to 999, thereby warning users the true SE cannot be estimated in this case. Similar events also occur in other applications of the Observed Fisher Information approach. The usual solution is to increase the sample size, and in our case to collect more studies for a larger $k$. 

There are extreme mixed sample size situations where the number of studies $k$ is small and the size of one study is much larger than that of others. Does one large study compensate for a small $k$? We conducted simulations with $k$ values being 10 or 20, and all but the largest $n_i$ being between 100 and 200. The largest $n_i$ is set to 100 times larger than other $n_i$'s. The results, included in Appendix~\ref{appendix:extreme}, show no significant difference compared to previous results for scenarios where all $n_i$'s are of similar size. This shows that the influence of an individual study size on the estimation accuracy is much smaller than the number of studies $k$, adding to the importance of having a larger $k$.

\subsection{Key Findings from Simulation Results}

Due to space limitation, we presented only a part of our simulation results. We now summarize the key findings of our simulation study based on all results. 

\begin{enumerate}[label=(\arabic*)]
    \item The proposed estimator $\hat{p}_{11}$ is accurate when the total number of studies is not too small ($k\geq 20$), regardless of the individual study size $n_i$. It is also robust against extreme variations in study sizes $n_i$.
    \item While a larger $k$ and larger $n_i$ will both give more accurate estimates, the impact of $k$ on accuracy is much greater than that of $n_i$. In a designed experiment situation where the total sample size $N=\sum_{i=1}^k n_i$ is fixed due to budget constraints but $k$ and individual $n_i$ can be controlled, we recommend a higher $k$ value and lower $n_i$ values for better estimation accuracy. 
    \item The Observed Fisher Information based estimator for the standard error of $\hat{p}_{11}$ is reliable, provided $k$ is not too small. Both normal approximation based confidence interval (\( CI_1 \)) and likelihood ratio confidence interval (\( CI_2 \)) work well, but in most cases the latter is superior with better accuracy and shorter length.
\end{enumerate}


\section{Real Data Analysis}
We now apply the proposed method to estimate a $p_{11}$ using summary-level data of 12 real-world datasets (see Table \ref{real_data}) sourced from \textbf{\small Project Data Sphere}, a platform that promotes data sharing for cancer research. Detailed discussions on dataset selection, data pre-processing, and dataset characteristics may be found in Appendix~\ref{appendix:realdata}.
The original datasets contain patient-level data with which we estimated $p_{11}$ directly using the corresponding sample proportion. Note that patient-level data and thus such a direct estimate of $p_{11}$ are not available in the privacy-constrained setting of this paper. The reason we choose these 12 datasets is that we can estimate $p_{11}$ using our method and also compare it with the direct estimate which is taken as the ``true value".

\begin{table*}[ht]
    \centering
    \caption{Summary of clinical study data. The columns represent the study ID, type, geographic region, and various sample statistics (e.g., abnormal BMI and White race counts).}
    \label{real_data}
    \vskip 0.15in
    \begin{small}
    \begin{sc}
    \begin{tabular}{ccccccc}
        \toprule
        \textbf{k} & \textbf{Study ID} & \textbf{Study} & \textbf{Region} & \textbf{Sample Size} & \makecell{\textbf{BMI=1} \\ \textbf{(Abnormal)}} & \makecell{\textbf{Race=1} \\ \textbf{(White)}} \\
        \midrule
        1  & NCT00081796  & Breast Cancer     & Northeast & 338   & 237   & 218   \\
        2  & NCT00081796  & Breast Cancer     & Midwest   & 377   & 268   & 302   \\
        3  & NCT00081796  & Breast Cancer     & South     & 663   & 459   & 391   \\
        4  & NCT00081796  & Breast Cancer     & West      & 393   & 235   & 222   \\
        5  & NCT00554229  & Prostate Cancer   & Northeast & 259   & 199   & 171   \\
        6  & NCT00554229  & Prostate Cancer   & Midwest   & 273   & 209   & 220   \\
        7  & NCT00554229  & Prostate Cancer   & South     & 512   & 393   & 288   \\
        8  & NCT00554229  & Prostate Cancer   & West      & 263   & 173   & 175   \\
        9  & NCT00981058  & Lung Cancer       &           & 545   & 280   & 455   \\
        10 & NCT01169259  & Cancer Prevention &           & 24193 & 16890 & 17662 \\
        11 & NCT00626548  & Prostate Cancer   &           & 652   & 457   & 463   \\
        12 & NCT00617669  & Prostate Cancer   &           & 426   & 296   & 331   \\
        \bottomrule
    \end{tabular}
    \end{sc}
    \end{small}
    \vskip -0.1in
\end{table*}

\begin{figure}[ht]
    \centering
    \includegraphics[width=\linewidth]{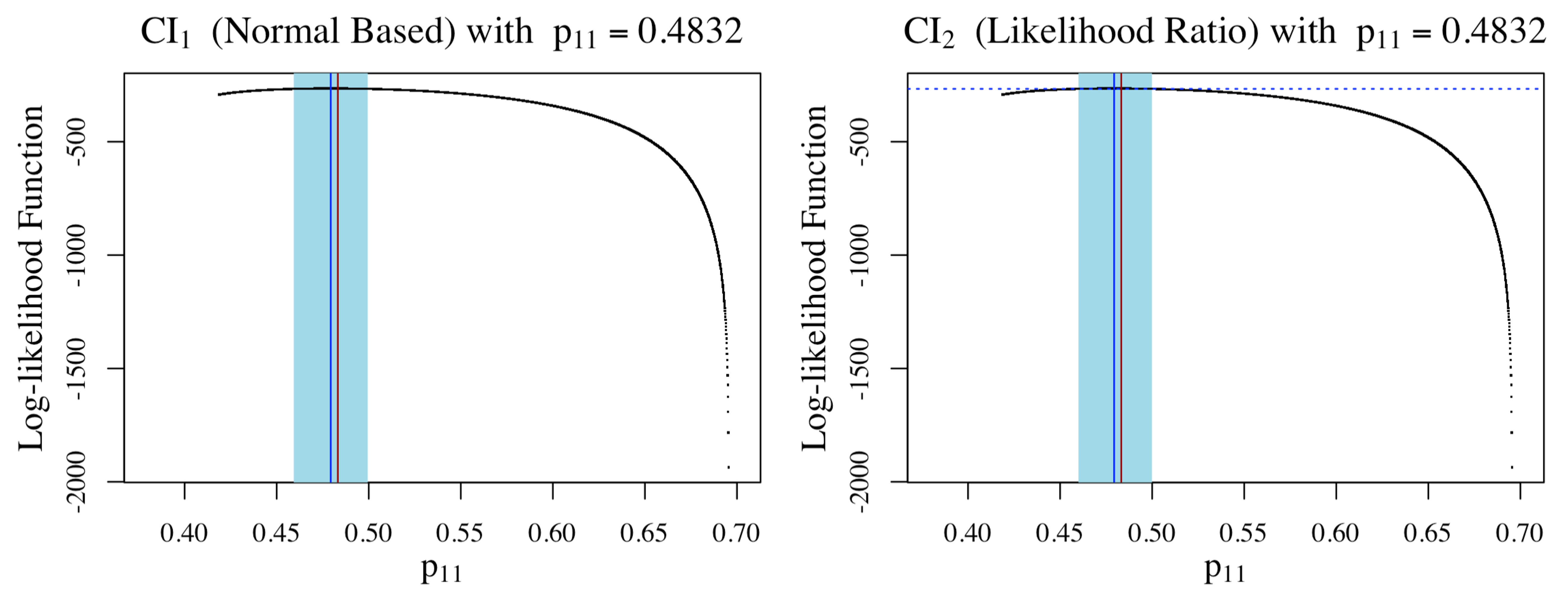}
    \caption{Point and interval estimation of $p_{11}$ for variables BMI and Race in Table \ref{real_data}. The red line represents the true $p_{11}$ ($0.4832$), the blue line represents the estimated value $\hat{p}_{11}$ (0.4794), and the blue-band indicates the $95\%$ confidence interval of ${p}_{11}$.}
    \label{application_CI}
\end{figure}
The estimate based on the summary-level data is $\hat{p}_{11}=0.4794$, very close to the true value $p_{11}=0.4832$. Fig. \ref{application_CI} also shows $p_{11}$ is in the $95\%$ confidence intervals. These demonstrate that the proposed method can provide accurate estimates even with a relatively small sample size ($k=12$).


\section{Concluding Remarks}

Simulation studies are an essential tool for designing clinical trials \cite{ref8}. However, a common challenge in such simulations is the generation of unrealistic data, especially when correlations between variables are not adequately considered. This challenge is exacerbated when only study-level data is available, a situation often encountered due to privacy concerns or proprietary interests. 

We developed a maximum likelihood-based method for estimating the joint distribution of two binary variables using only marginal summary data. The highlight of this work is the maximum likelihood estimator and its associated standard error estimator for the key parameter $p_{11}$, which is critical in characterizing the correlation between the two variables. By accurately estimating the $p_{11}$, our method effectively mitigates the risk of generating unrealistic data, which is crucial in clinical trial simulations where data validity can significantly impact trial design and outcomes. The robustness of our method has been demonstrated through extensive simulation studies under various scenarios, and through its application to real-world cancer-related datasets. These results show that our approach not only improves the accuracy of correlation estimation but also enhances the overall reliability of the generated data, making it a valuable tool for researchers working with limited data access.

Although our initial motivation was high-quality simulations in clinical trial design, the proposed method has potential applications in broader fields facing similar challenges, such as the Internet of Things and privacy-sensitive distributed environments. As these domains often deal with incomplete or summary-level data, our method serves as a valuable new tool to effectively capture variable dependency while ensuring privacy.

As the clinical trial simulation is a multidisciplinary effort, it requires clinical pharmacologists, pharmacokineticists, statisticians, preclinical pharmacologists, and other key stakeholders to come together to discuss the known and unknown information about drugs and the connections between this information \cite{ref9}. In future research, we plan to extend our approach to accommodate multiple variables of potentially mixed types (e.g., continuous, binary, or thresholded), thereby supporting more general scenarios for estimating interactions. However, as the number of variables increases, so does the challenge of estimating the joint distribution, which may render the task theoretically or computationally infeasible. In such cases, dimensionality reduction techniques or structured probabilistic models may provide practical solutions. Overall, we believe that statistical inference based solely on marginal summary data could form a new paradigm in privacy-sensitive environments, offering a scalable and privacy-preserving framework for learning variable interactions.


\bibliographystyle{plain}
\bibliography{ref}

\begin{thebibliography}{10}

\bibitem{ref9}
Peter~L. Bonate.
\newblock Clinical trial simulation in drug development.
\newblock {\em Pharmaceutical Research}, 17(3):252--256, March 2000.

\bibitem{ref8}
Maria~Mori Brooks, Al~Hallstrom, and Monika Peckova.
\newblock A simulation study used to design the sequential monitoring plan for a clinical trial.
\newblock {\em Statistics in Medicine}, 14(20):2227--2237, 1995.

\bibitem{nref1}
Joseph~A. DiMasi, Henry~G. Grabowski, and Ronald~W. Hansen.
\newblock The cost of drug development.
\newblock {\em New England Journal of Medicine}, 372(20):1972--1972, 2015.

\bibitem{FDA2019}
U.S. Food and Drug Administration.
\newblock Adaptive designs for clinical trials of drugs and biologics: Guidance for industry, 2019.
\newblock Available: \url{https://www.fda.gov/media/78495/download}.

\bibitem{FDA2022}
U.S. Food and Drug Administration.
\newblock Complex innovative trial design meeting program, 2022.
\newblock Available: \url{https://www.fda.gov/drugs/development-resources/complex-innovative-trial-design-meeting-program}.

\bibitem{ref2}
M.~A. Hamdan and E.~O. Martinson.
\newblock Maximum likelihood estimation in the bivariate binomial (0,1) distribution:application {TO} 2×2 tables.
\newblock {\em Australian Journal of Statistics}, 13(3):154--158.

\bibitem{ref1}
M.~A. Hamdan and M.~O Nasro.
\newblock Maximum likelihood estimation of the parameters of the bivariate binomial distribution.
\newblock {\em Communications in Statistics - Theory and Methods}, 15(3):747--754, 1986.

\bibitem{nref2}
Richard~K. Harrison.
\newblock Phase ii and phase iii failures: 2013–2015.
\newblock {\em Nature Reviews Drug Discovery}, 15(12):817--818, 2016.

\bibitem{nref3}
Michael Hay, David~W. Thomas, John~L. Craighead, Celia Economides, and Jesse Rosenthal.
\newblock Clinical development success rates for investigational drugs.
\newblock {\em Nature Biotechnology}, 32(1):40--51, 2014.

\bibitem{ref10}
N~Holford, S~C Ma, and B~A Ploeger.
\newblock Clinical trial simulation: A review.
\newblock {\em Clinical Pharmacology \& Therapeutics}, 88(2):166--182, 2010.

\bibitem{Holford2010}
N~Holford, S~C Ma, and B~A Ploeger.
\newblock {Clinical Trial Simulation: A Review}.
\newblock {\em Clinical Pharmacology \& Therapeutics}, 88(2):166--182, 2010.

\bibitem{Mayer2019}
Cristiana Mayer, Inna Perevozskaya, Sergei Leonov, Vladimir Dragalin, Yili Pritchett, Alun Bedding, Alan Hartford, Parvin Fardipour, and Greg Cicconetti.
\newblock {Simulation Practices for Adaptive Trial Designs in Drug and Device Development}.
\newblock {\em Statistics in Biopharmaceutical Research}, 11(4):325--335, 2019.

\bibitem{Walker2020}
Martin Walker, Jonathan I.~D. Hamley, Philip Milton, Frédéric Monnot, Belén Pedrique, and Maria-Gloria Basáñez.
\newblock {Designing antifilarial drug trials using clinical trial simulators}.
\newblock {\em Nature Communications}, 11(1):2685, 2020.

\bibitem{ref14}
Wei Wang, Ying Ma, Yangxin Huang, and Henian Chen.
\newblock Generalizability analysis for clinical trials: a simulation study.
\newblock {\em Statistics in Medicine}, 36(10):1523--1531, 2017.

\bibitem{Westfall2008}
Peter~H. Westfall, Kuenhi Tsai, Stephan Ogenstad, Alin Tomoiaga, Scott Moseley, and Yonggang Lu.
\newblock {Clinical Trials Simulation: A Statistical Approach}.
\newblock {\em Journal of Biopharmaceutical Statistics}, 18(4):611--630, 2008.

\bibitem{D4}
Fadila Zerka, Samir Barakat, Sean Walsh, Marta Bogowicz, Ralph T.~H. Leijenaar, Arthur Jochems, Benjamin Miraglio, David Townend, and Philippe Lambin.
\newblock Systematic review of privacy-preserving distributed machine learning from federated databases in health care.
\newblock {\em JCO Clinical Cancer Informatics}, (4):184--200, 2020.
\newblock PMID: 32134684.

\end{thebibliography}


\newpage
\appendix
\onecolumn
\section{Detailed Derivation of Maximum Likelihood Estimation}
\label{appendix:MLE}

This section provides a detailed derivation of the Maximum Likelihood Estimation (MLE) for the parameters of the bivariate binomial distribution based on the joint PMF \eqref{pmf}. Specifically, we start by revisiting the joint pmf and the log-likelihood function. We then compute the partial derivatives of the log-likelihood function with respect to the parameters \(p_1\), \(p_2\), and \(p_{11}\). Finally, the closed-form MLEs for \(p_1\) and \(p_2\) are obtained by solving the resulting system of equations.

\subsection{Joint Distribution and Likelihood Function}

From Chapter 2, the joint distribution and the corresponding log-likelihood function for the model are defined as follows.

\textbf{Joint Distribution}
The joint PMF for the observed data is given by:
$$b(x_i, y_i, n_i; \bm{p}) = \sum_{z_i=0}^{\min(x_i, y_i)} T(x_i, y_i, z_i, n_i; \bm{p}),$$

where $\bm{p} = (p_1, p_2, p_{11}),$
$$T(x_i, y_i, z_i, n_i; \bm{p}) = \frac{n_i! \cdot p_{11}^{z_i} (p_1 - p_{11})^{x_i - z_i} (p_2 - p_{11})^{y_i - z_i} (1 - p_1 - p_2 + p_{11})^{n_i - x_i - y_i + z_i}}{z_i! (x_i - z_i)! (y_i - z_i)! (n_i - x_i - y_i + z_i)!}.$$

\textbf{Log-Likelihood Function}

The log-likelihood function is obtained by summing the log-transformed joint pmf over all observations:

$$l(\bm{p})=\sum_{i=1}^k \log b(x_i,y_i,n_i;\bm{p}).$$

To estimate \(\bm{p}\), we derive the partial derivatives of the log-likelihood function with respect to each parameter \(p_1\), \(p_2\), and \(p_{11}\), and then solve the resulting system of equations. The partial derivatives are given by:

$$\frac{\partial l(\bm{p})}{\partial \bm{p}} = 
\sum_{i=1}^{k} \frac{\partial \ln b(x_i, y_i, n_i;\bm{p})}{\partial \bm{p}} = 
\sum_{i=1}^{k} \frac{\frac{\partial b(x_i, y_i, n_i;\bm{p})}{\partial \bm{p}}}{b(x_i, y_i, n_i;\bm{p})} = 
\sum_{i=1}^{k} \frac{\sum_{z_i}\frac{\partial T(x_i, y_i, z_i, n_i;\bm{p})}{\partial \bm{p}}}{\sum_{z_i} T(x_i, y_i, z_i, n_i;\bm{p})}.$$

By setting these derivatives to zero, we derive the MLEs for the parameters \(p_1\), \(p_2\), and \(p_{11}\).

\subsection{Partial Derivative with Respect to $p_1$ and $p_2$}

To compute the MLEs of the parameters, we first derive the partial derivatives of the log-likelihood function with respect to \(p_1\) and \(p_2\). The partial derivative with respect to \(p_1\) is expressed as:

$$\frac{\partial l(\bm{p})}{\partial p_1} = 
\sum_{i=1}^{k} \frac{\partial \ln b(x_i, y_i, n_i;\bm{p})}{\partial p_1} = 
\sum_{i=1}^{k} \frac{\frac{\partial b(x_i, y_i, n_i;\bm{p})}{\partial p_1}}{b(x_i, y_i, n_i;\bm{p})}
= 
\sum_{i=1}^{k} \frac{\sum_{z_i}\frac{\partial T(x_i, y_i, z_i, n_i;\bm{p})}{\partial p_1}}{\sum_{z_i} T(x_i, y_i, z_i, n_i;\bm{p})}.$$

Here, the numerator involves the derivative of \(T(x_i, y_i, z_i, n_i; \bm{p})\) with respect to \(p_1\). This is given as:

$$\sum_{z_i}\frac{\partial T(x_i, y_i, z_i, n_i;\bm{p})}{\partial p_1}
$$

$$= \sum_{z_i}\frac{n_i! \cdot p_{11}^{z_i}(p_2-p_{11})^{y_i-z_i}}
{z_i!(x_i-z_i)!(y_i-z_i)!(n_i-x_i-y_i+z_i)!}
\cdot
\frac{\partial [(p_1-p_{11})^{x_i-z_i}(1-p_1-p_2+p_{11})^{n_i-x_i-y_i+z_i}]}{\partial p_1}$$

$$=\sum_{z_i}T(x_i,y_i,z_i,n_i)
\frac{[(x_i-z_i)(1-p_2)+(y_i-n_i)(p_1-p_{11})]}{(p_1-p_{11})(1-p_1-p_2+p_{11})}$$

$$=\frac{(1-p_2)\sum_{z_i}T(x_i,y_i,z_i,n_i) (x_i-z_i)}
{(p_1-p_{11})(1-p_1-p_2+p_{11})}
+\frac{\sum_{z_i}T(x_i,y_i,z_i,n_i)(y_i-n_i)}
{(1-p_1-p_2+p_{11})}.$$

Divided by \(\sum_{z_i} T(x_i, y_i, z_i, n_i; \bm{p})\), we obtain:

$$\frac{\sum_{z_i}\frac{\partial T(x_i, y_i, z_i, n_i;\bm{p})}{\partial p_1}}{\sum_{z_i} T(x_i, y_i, z_i, n_i;\bm{p})}
=
\frac{(1-p_2)(x_i-\frac{\sum_{z_i} T(x_i,y_i,z_i,n_i) \cdot z_i}{\sum_{z_i} T(x_i,y_i,z_i,n_i)})}{(p_1-p_{11})(1-p_1-p_2+p_{11})}
+\frac{(y_i-n_i)}{(1-p_1-p_2+p_{11})}.
$$

Summing over all \(i\), the resulting equation is as follows:

$$
\sum_{i=1}^{k} \frac{\sum_{z_i}\frac{\partial T(x_i, y_i, z_i, n_i;\bm{p})}{\partial p_1}}{\sum_{z_i} T(x_i, y_i, z_i, n_i;\bm{p})}
= 0 \iff
$$

$$
\sum_{i=1}^k\left\{\frac{(1-p_2)}{(p_1-p_{11})}x_i+y_i-n_i-\frac{(1-p_2)(\sum_{z_i}z_i \cdot T(x_i,y_i,z_i,n_i))}
{(p_1-p_{11})(\sum_{z_i}T(x_i,y_i,z_i,n_i))}\right\}
= 0.
$$

Finally, this gives the equation for \(\frac{\partial l(\bm{p})}{\partial p_1}\):

\begin{equation}
    \label{seq1}
    \frac{\partial l(\bm{p})}{\partial p_{1}} = 
    \frac{1-p_2}{p_1-p_{11}}\sum_ix_i+\sum_iy_i-\sum_in_i-\frac{1-p_2}{p_1-p_{11}} \cdot S=0,
\end{equation}
where
$$S(x_i, y_i, n_i; \bm{p})
=
\sum_{i=1}^k\frac{\sum_{z_i}z_i \cdot T(x_i,y_i,z_i,n_i)}{b(x_i,y_i,n_i)}.$$

By symmetry, the partial derivative with respect to \(p_2\) can be derived analogously, resulting in:

\begin{equation}
    \label{seq2}
    \frac{\partial l(\bm{p})}{\partial p_{2}} = 
    \frac{1-p_1}{p_2-p_{11}}\sum_iy_i+\sum_ix_i-\sum_in_i-\frac{1-p_1}{p_2-p_{11}} \cdot S=0.
\end{equation}

\subsection{Partial Derivative with Respect to $p_{11}$}

To compute the partial derivative of the log-likelihood function with respect to \(p_{11}\), we start with the following expression:

$$\frac{\partial l(\bm{p})}{\partial p_{11}} = 
\sum_{i=1}^{k} \frac{\partial \ln b(x_i, y_i, n_i;\bm{p})}{\partial p_{11}} = 
\sum_{i=1}^{k} \frac{\frac{\partial b(x_i, y_i, n_i;\bm{p})}{\partial p_{11}}}{b(x_i, y_i, n_i;\bm{p})}
= 
\sum_{i=1}^{k} \frac{\sum_{z_i}\frac{\partial T(x_i, y_i, z_i, n_i;\bm{p})}{\partial p_{11}}}{\sum_{z_i} T(x_i, y_i, z_i, n_i;\bm{p})}.$$

Here, the numerator contains the derivative of \(T(x_i, y_i, z_i, n_i; \bm{p})\) with respect to \(p_{11}\). By applying the definition of \(T(x_i, y_i, z_i, n_i; \bm{p})\), we expand it as:

$$\sum_{z_i}\frac{\partial T(x_i, y_i, z_i, n_i;\bm{p})}{\partial p_{11}}
$$

$$= \sum_{z_i}\frac{n_i!}
{z_i!(x_i-z_i)!(y_i-z_i)!(n_i-x_i-y_i+z_i)!}
\cdot
$$

$$ \frac{\partial \left[
p_{11}^{z_i}
(p_1-p_{11})^{x_i-z_i}(p_2-p_{11})^{y_i-z_i}(1-p_1-p_2+p_{11})^{n_i-x_i-y_i+z_i}
\right]}{\partial p_{11}}
$$

$$= \sum_{z_i}T(x_i,y_i,z_i,n_i) \cdot 
\left[\frac{z_i}{p_{11}}-\frac{(x_i-z_i)}{(p_1-p_{11})}
-\frac{(y_i-z_i)}{(p_2-p_{11})}+\frac{(n_i-x_i-y_i+z_i)}{(1-p_1-p_2+p_{11})}
\right]$$

$$
=\sum_{z_i}T(x_i,y_i,z_i,n_i) \cdot 
\left[-\frac{x_i(1-p_2)}{(p_1-p_{11})(1-p_1-p_2+p_{11})} 
-\frac{y_i(1-p_1)}{(p_2-p_{11})(1-p_1-p_2+p_{11})} 
\right.
$$

$$
+\frac{n_i}{(1-p_1-p_2+p_{11})}
+ z_i \left(\frac{1}{p_{11}} - \frac{1}{p_1-p_{11}} - \frac{1}{p_2-p_{11}} 
+ \frac{1}{1-p_1-p_2+p_{11}}\right)\Bigg].
$$

Divided by the denominator \(\sum_{z_i} T(x_i, y_i, z_i, n_i; \bm{p})\), we obtain:

$$
\sum_{i=1}^{k} \frac{\sum_{z_i}\frac{\partial T(x_i, y_i, z_i, n_i;\bm{p})}{\partial p_{11}}}{\sum_{z_i} T(x_i, y_i, z_i, n_i;\bm{p})} = 0 \iff
$$

$$
-\frac{1-p_2}{p_1-p_{11}}\sum_ix_i - \frac{1-p_1}{p_2-p_{11}}\sum_iy_i + \sum_in_i 
$$
$$
+ \left(\frac{1}{p_{11}} - \frac{1}{p_1-p_{11}} - \frac{1}{p_2-p_{11}} 
+ \frac{1}{1-p_1-p_2+p_{11}}\right)(1-p_1-p_2+p_{11})S = 0,
$$

where $$S(x_i, y_i, n_i;\bm{p})=\sum_{i=1}^k\frac{\sum_{z_i}z_i \cdot T(x_i,y_i,z_i,n_i)}{b(x_i,y_i,n_i)}.$$

Substituting Eq.~\ref{seq1} and Eq.~\ref{seq2} into the equation above, we have:

$$\sum_ix_i+\sum_iy_i-\sum_in_i+(\frac{1-p_1-p_2-p_{11}}{p_{11}}
+\frac{p_{11}-p_1}{p_1-p{11}}+\frac{p_{11}-p_2}{p_2-p_{11}}+1) 
\cdot S=0$$

Thus, the partial derivative with respect to \(p_{11}\) is given by:
$$\frac{\partial l(\bm{p})}{\partial p_{11}} = 
\sum_ix_i+\sum_iy_i-\sum_in_i+\frac{1-p_1-p_2}{p_{11}}\cdot S=0,$$
where $$S(x_i, y_i, n_i;\bm{p})=\sum_{i=1}^k\frac{\sum_{z_i}z_i \cdot T(x_i,y_i,z_i,n_i)}{b(x_i,y_i,n_i)}.$$

\subsection{Maximum Likelihood Estimation of \(p_1\) and \(p_2\)}
To summarize the derivation, we have the following system of equations:

\begin{subequations}\label{eq_s}
\begin{align}
\frac{\partial l(\bm{p})}{\partial p_1} &= \frac{1 - p_2}{p_1 - p_{11}} \sum_i x_i + \sum_i y_i - \sum_i n_i - \frac{1 - p_2}{p_1 - p_{11}} S = 0 \label{eq_s1}, \\
\frac{\partial l(\bm{p})}{\partial p_2} &= \frac{1 - p_1}{p_2 - p_{11}} \sum_i y_i + \sum_i x_i - \sum_i n_i - \frac{1 - p_1}{p_2 - p_{11}} S = 0 \label{eq_s2}, \\
\frac{\partial l(\bm{p})}{\partial p_{11}} &= \sum_i x_i + \sum_i y_i - \sum_i n_i + \frac{1 - p_1 - p_2}{p_{11}} S = 0 \label{eq_s3},
\end{align}
\end{subequations}

where 
$S(x_i, y_i, n_i; \bm{p}) = \sum_{i=1}^k \sum_{z_i} \frac{z_i \cdot T(x_i, y_i, z_i, n_i;\bm{p})}{b(x_i, y_i, n_i;\bm{p})}$.

To solve this system of equations, we begin by manipulating the equations to isolate \(S\). First, subtract Eq.~\ref{eq_s3} from Eq.~\ref{eq_s1}, yielding:

$$
\left( \frac{1-p_2}{p_1-p_{11}}-1 \right)\sum_i x_i = 
\left( \frac{1-p_2}{p_1-p_{11}} + \frac{1-p_1-p_2}{p_{11}} \right)\ S
.$$

Simplifying this equation leads to 

\begin{equation}
\label{seq_mid1}
S = \frac{p_{11}}{p_1} \sum_i x_i.
\end{equation}

Similarly, subtracting Eq.~\ref{eq_s3} from Eq.~\ref{eq_s2} and simplifying gives:

\begin{equation}
\label{seq_mid2}
S = \frac{p_{11}}{p_2} \sum_i y_i.
\end{equation}

Dividing Eq.~\ref{seq_mid2} by Eq.~\ref{seq_mid1} yields a ratio between \(p_1\) and \(p_2\):

\begin{equation}
\label{seq_mid3}
\frac{p_1}{p_2} = \frac{\sum_i x_i}{\sum_i y_i}.
\end{equation}

Substituting Eq.~\ref{seq_mid3} and Eq.~\ref{seq_mid1} into Eq.~\ref{eq_s1}, we obtain the MLE of \(p_1\):

\begin{equation}
\label{seq_mle1}
\hat{p}_1 = \frac{\sum_i x_i}{\sum_i n_i}.
\end{equation}

Similarly, substituting Eq.~\ref{seq_mid3} and Eq.~\ref{seq_mid2} into Eq.~\ref{eq_s2} gives the MLE of \(p_2\):

\begin{equation}
\label{seq_mle1}
\hat{p}_2 = \frac{\sum_i y_i}{\sum_i n_i}.
\end{equation}

\section{Normality Assessment of $p_{11}$ Estimation Results}
\label{appendix:normal}

This section provides a detailed assessment of the normality of the \(\hat{p}_{11}\) results obtained from simulations, as described in Section 3.3 of the main paper. To validate the approximate normality of the $\hat{p}_{11}$, we present histograms and QQ plots for $\hat{p}_{11}$ based on 1000 repetitions of the experiment conducted across 20 different settings. 

\begin{figure}[H]
    \centering
    \begin{minipage}{\textwidth}
        \centering
        \includegraphics[width=14cm]{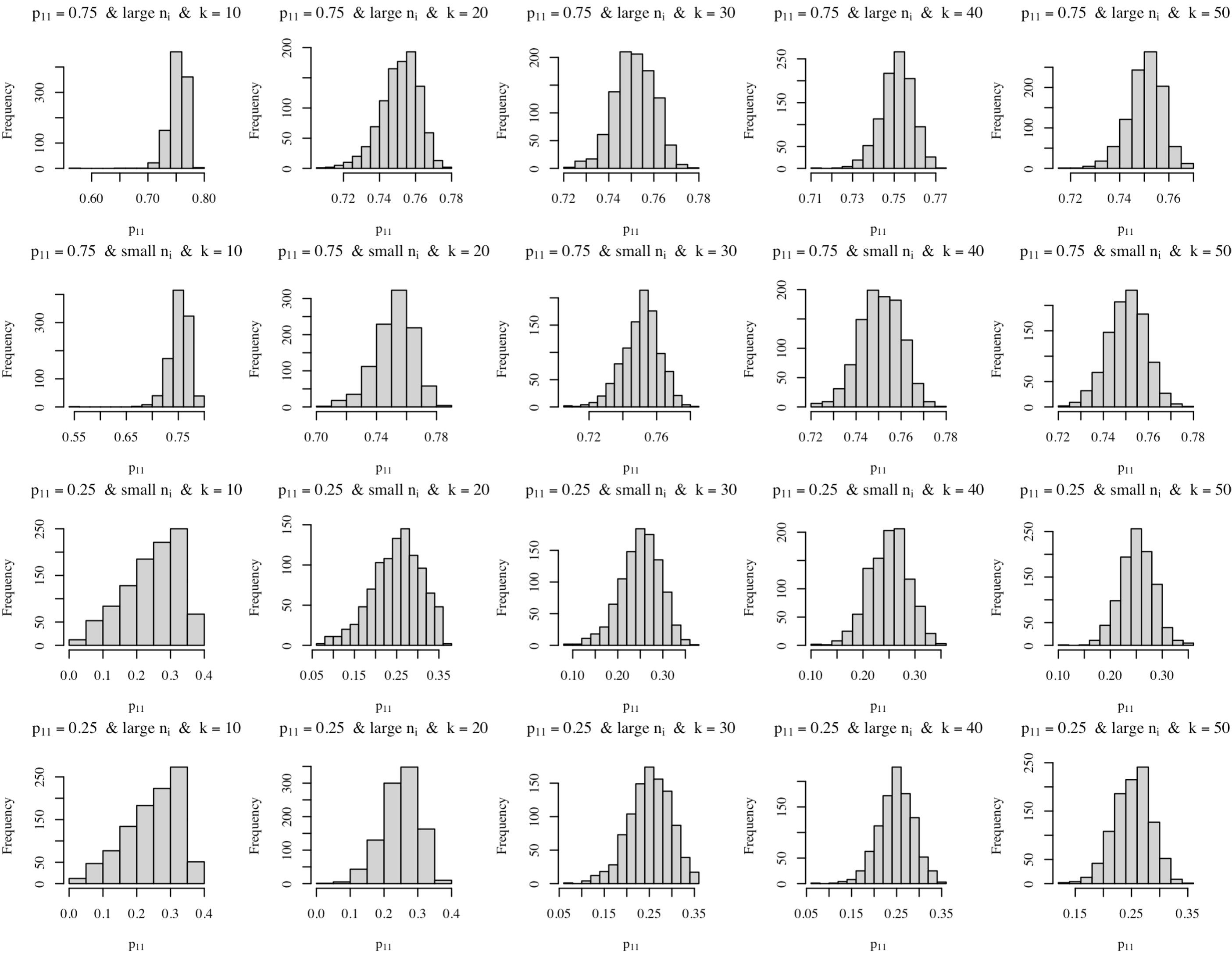} 
    \end{minipage}
    
    \caption{Histograms of the estimated \(p_{11}\) obtained from 1000 repetitions of the experiment conducted under 20 different settings. Each histogram corresponds to a specific experimental configuration, illustrating the distribution of \(p_{11}\) estimates.}
    \label{s_hist}
\end{figure}

\begin{figure}[H]
    \centering
    \begin{minipage}{\textwidth}
        \centering
        \includegraphics[width=14cm]{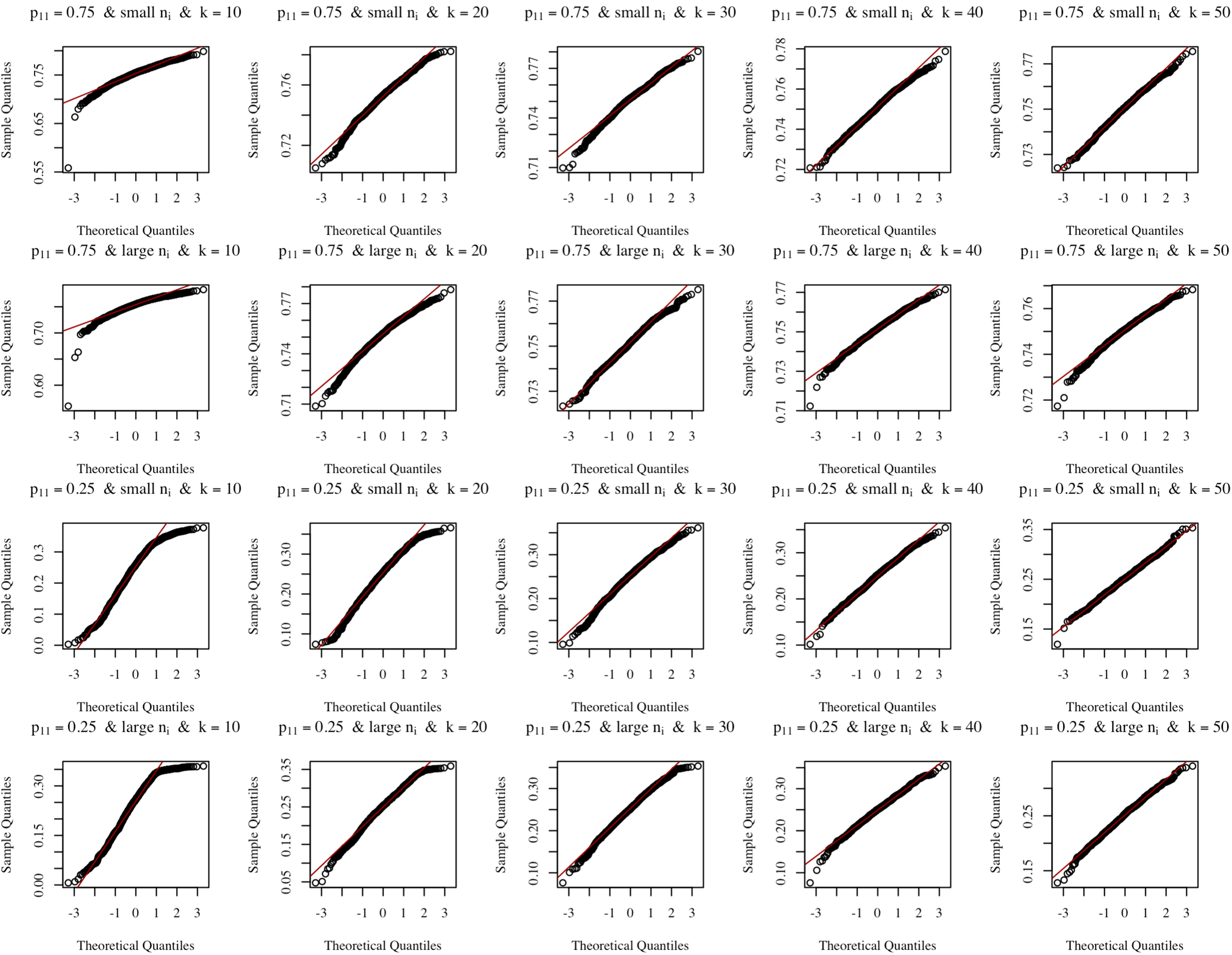} 
    \end{minipage}
    \caption{QQ plots comparing the quantiles of the estimated \(p_{11}\) distributions to the theoretical quantiles of a standard normal distribution. These plots validate the approximate normality of the \(p_{11}\) estimates, except in cases with small sample sizes (e.g., \(k=10\)).}
    \label{s_qq}
\end{figure}

The histograms (Figure \ref{s_hist}) depict the distribution of the $\hat{p}_{11}$ under various settings, illustrating the shape and spread of the estimates. Meanwhile, the QQ plots (Figure \ref{s_qq}) compare the quantiles of the empirical $\hat{p}_{11}$ distributions to those of a standard normal distribution, providing a visual check for normality.

It is observed that in most scenarios, the estimation results exhibit approximate normality. However, for \(k=10\), the results appear more concentrated, deviating from the normality assumption. This deviation is attributed to the insufficient sample size when \(k=10\), emphasizing the importance of conducting parameter estimation with larger sample sizes (e.g., \(k > 10\)) for more reliable and accurate results.

\begin{figure}[H]
    \centering
    \includegraphics[width=14cm]{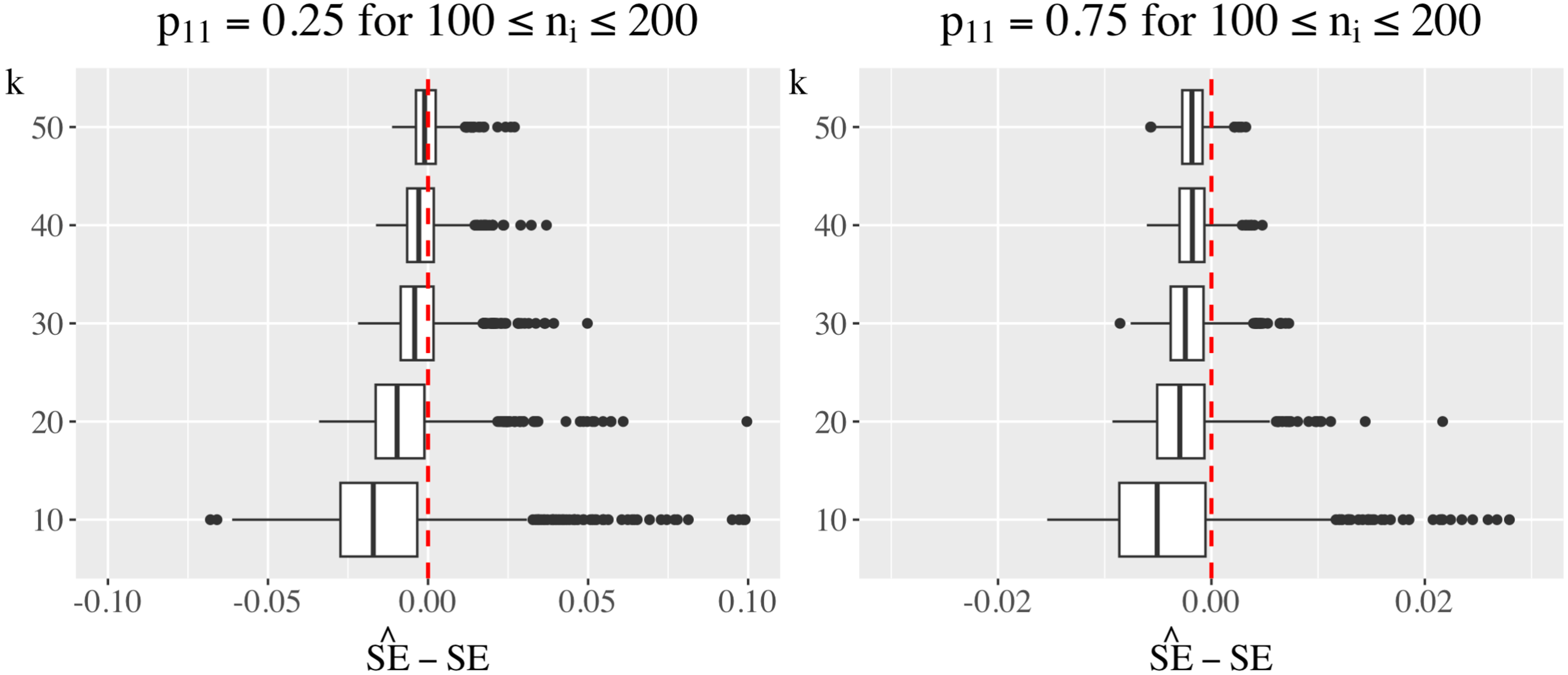}
    \caption{Box plots of bias ($\hat{SE}$ $-$ true SE) of the estimated standard error for 10 scenarios. The red line represents $y=0$.}
    \label{ab_diff_se}
\end{figure}

Figure \ref{ab_diff_se} illustrates the bias of the standard error (SE) across 1000 simulations, corresponding to an alternative scenario of Figure 1 in the main text (where the range of $n_i$ is 100–200). The patterns observed in both scenarios are similar, demonstrating that the bias of SE becomes increasingly accurate as $k$ increases.

\section{Impact of Sample Size \(n_i\) in an Extreme Scenario}
\label{appendix:extreme}

This section provides additional details on the simulation results under an extreme scenario, as discussed in Section 3.4 of the main text. In this scenario, the number of sample groups \(k\) is set to very small values (\(k=10\) and \(k=20\)), but one of the groups is assigned a significantly larger sample size \(n_i\) (100 times greater than the other \(n_i\) values, which range from 100 to 200). This setup is designed to evaluate the influence of an extremely large \(n_i\) on the estimation of \(p_{11}\) and the standard error.

\begin{figure}[H]
    \centering
    \includegraphics[width=14cm]{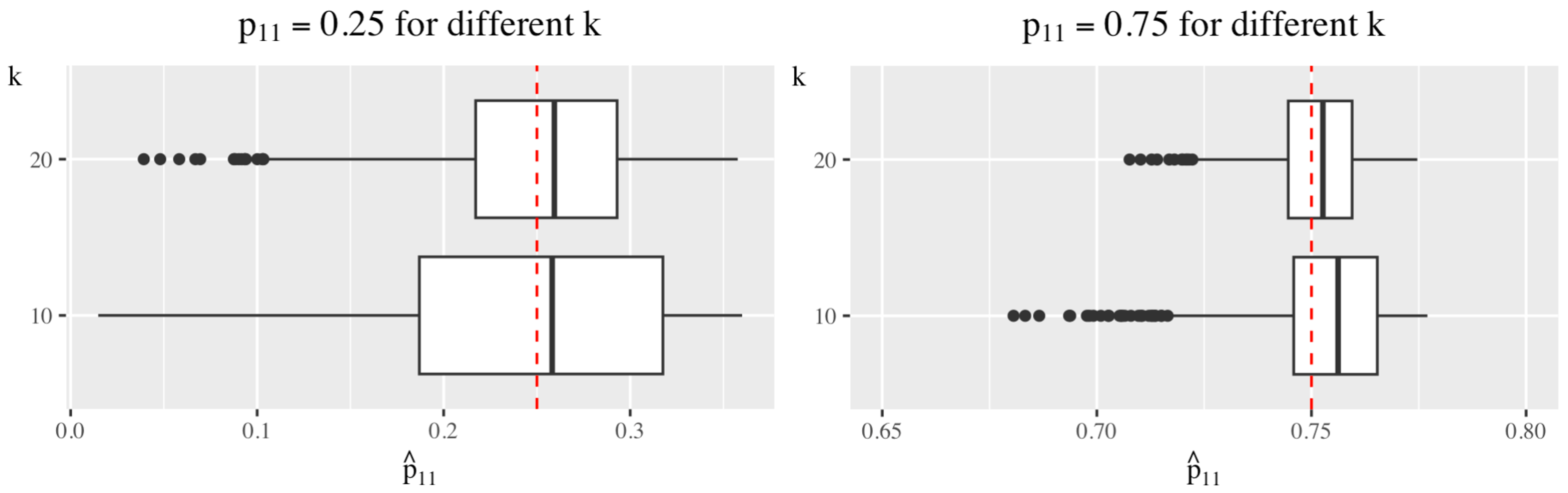}
    \caption{Box plots of the $\hat{p}_{11}$ results in the extreme scenario. The red lines represent the true values of $p_{11}$ ($0.25$ and $0.75$, respectively).}
    \label{extreme_p11_est}
\end{figure}

\begin{figure}[H]
    \centering
    \includegraphics[width=14cm]{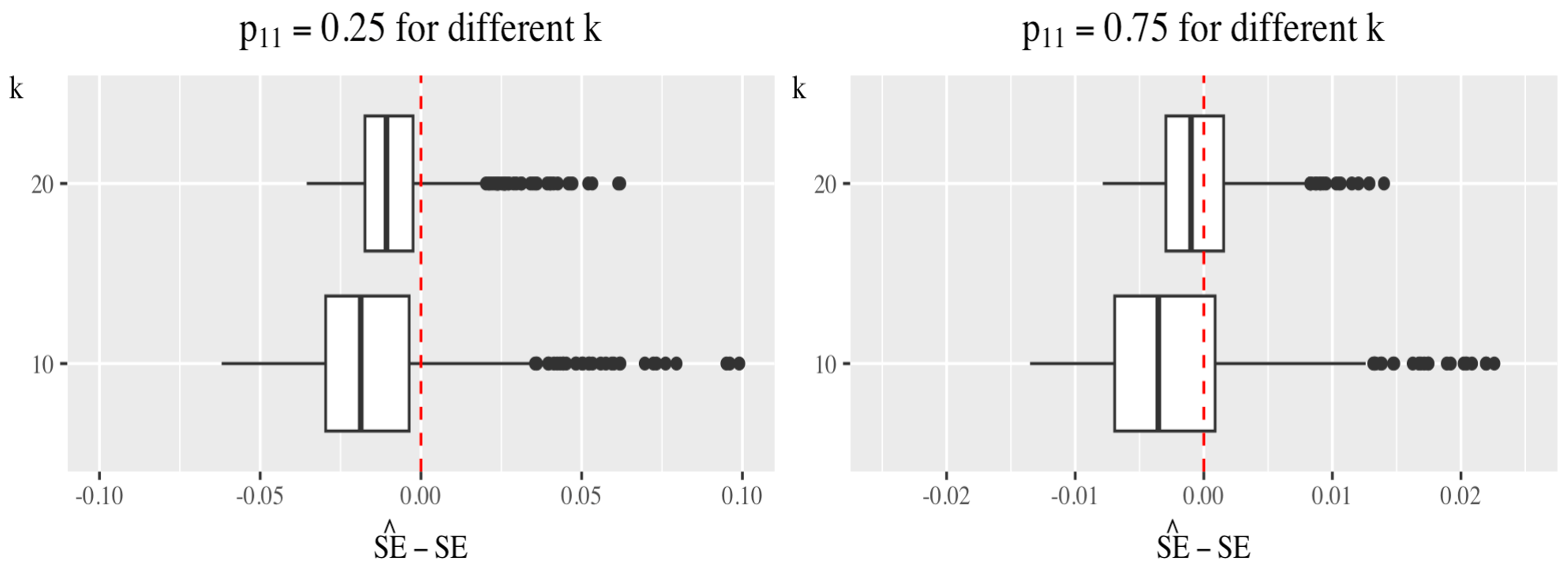}
    \caption{Box plots of bias ($\hat{SE}$ $-$ true SE) of the estimated standard error for the extreme scenario. The red line represents $y=0$.}
    \label{extreme_SE}
\end{figure}

\begin{figure}[H]
    \centering
    \begin{minipage}{\textwidth}
        \centering
        \includegraphics[width=12cm]{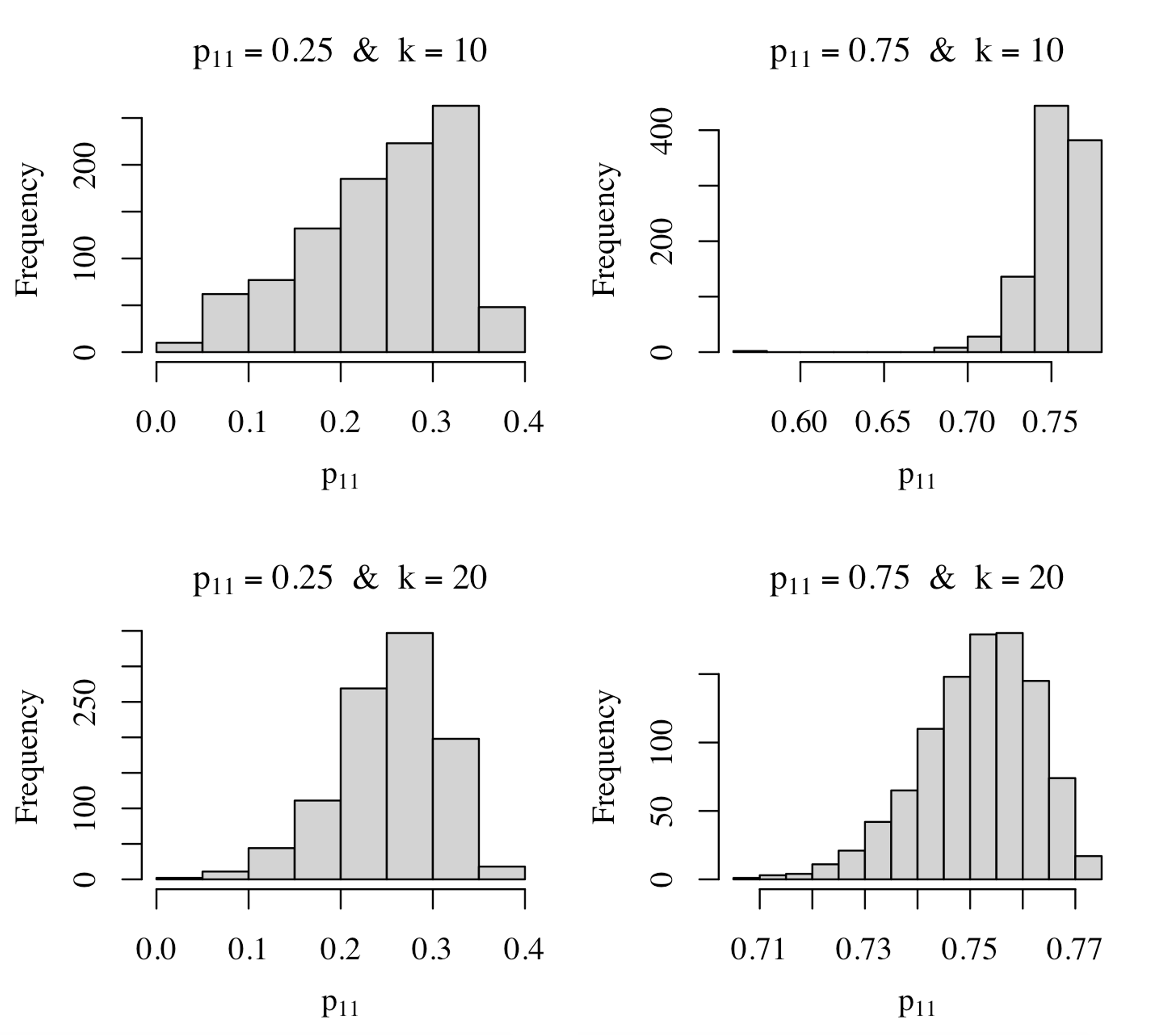} 
    \end{minipage}
    \caption{Histograms of $\hat{p}_{11}$ for the extreme scenario.}
    \label{extreme_hist}
\end{figure}

\begin{figure}[H]
    \centering
    \begin{minipage}{\textwidth}
        \centering
        \includegraphics[width=12cm]{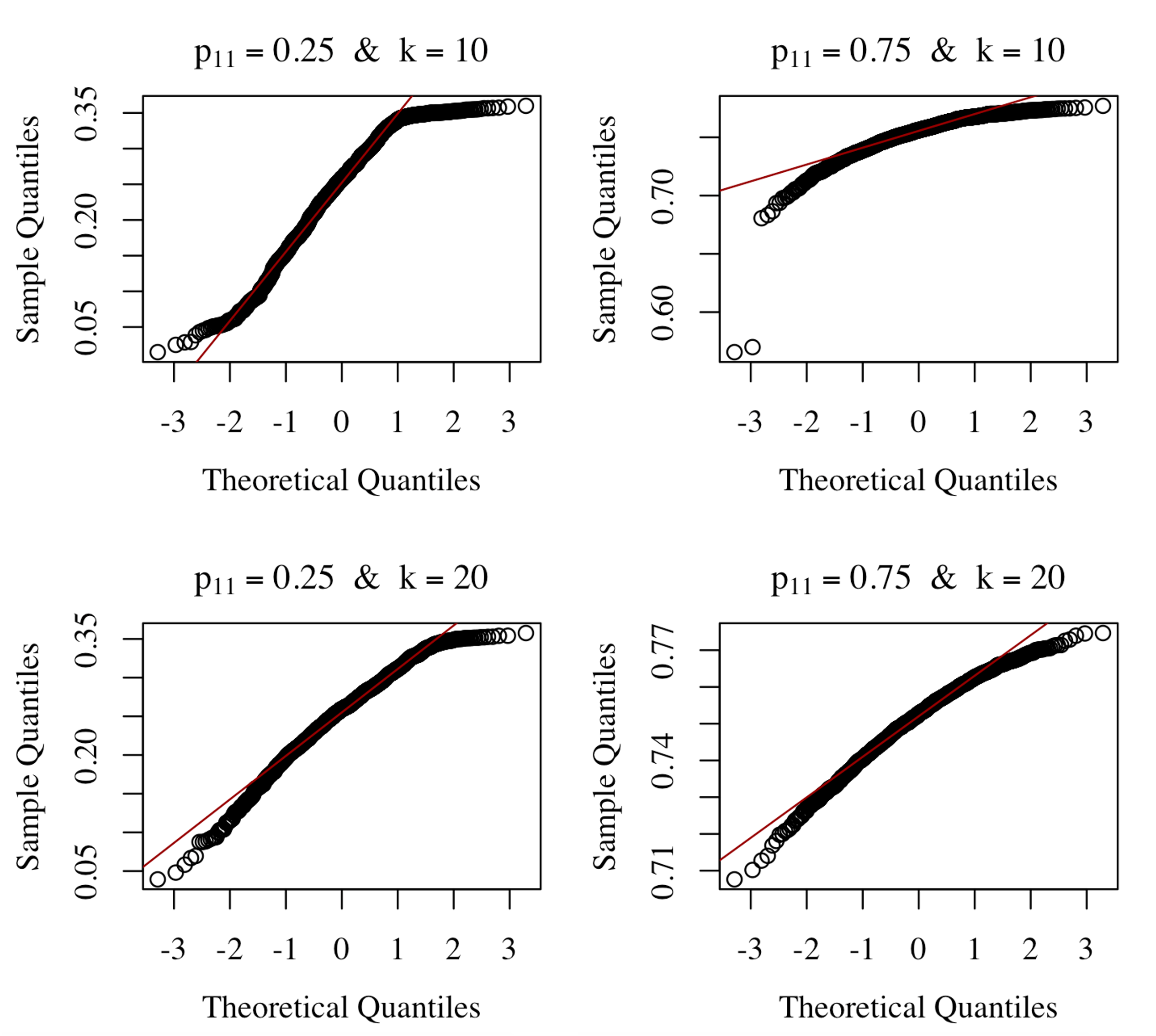} 
    \end{minipage}

    \caption{QQ plots of $\hat{p}_{11}$ for the extreme scenario.}
    \label{qq_extreme}
\end{figure}

The results, presented in Figures \ref{extreme_p11_est} to \ref{qq_extreme}, show no significant deviation from the findings of the previous simulations. Specifically, the estimation results of \(p_{11}\) remain consistent with the true values, and the bias in \(\hat{SE}\) is minimal. These findings suggest that the influence of \(n_i\) on the estimation is negligible compared to the number of sample groups (\(k\)), highlighting the importance of \(k\) in parameter estimation.

\section{Details on Real Data Analysis}
\label{appendix:realdata}

\subsection{Data Description and Selection Criteria}

The data for this study were sourced from \textbf{Project Data Sphere}, a platform that promotes data sharing for cancer research. We accessed patient-level data from 90 freely available cancer-related studies. Among these studies, many included various binary variables. We aimed to select common baseline variables, such as BMI (normal or abnormal), race (ethnic minority or majority), gender, and smoking status. These baseline characteristics (e.g., sex, age, severity of the disease, and racial groups) are primary factors in the generalizability of clinical trials~\cite{ref14}. We applied the following criteria to select datasets and binary variables for our empirical analysis:

\textbf{Consistency of Variables Across Studies:} The selected studies needed to include the same two binary variables. For example, studies on prostate cancer only involved male participants, making ``gender'' unavailable as a variable.\\
\textbf{Uniform Definition of Binary Variables:} The definition of binary variables had to be consistent across selected studies. While many studies included smoking-related variables, the descriptions varied significantly (e.g., recent smoking behavior vs. lifetime smoking history).\\
\textbf{Avoidance of Dataset Duplication:} To prevent redundancy, we ensured that no selected study used identical original datasets. During the review process, we identified that many of the 90 studies were based on overlapping datasets, likely due to the limited availability of publicly accessible patient-level data.

By applying these criteria, we ultimately selected 6 studies for analysis, focusing on BMI and RACE as the binary variables for our empirical analysis.

\subsection{Data Preprocessing}

For each dataset, we preprocessed the patient-level data as follows. We began by removing duplicate samples based on the unique ID assigned to each participant. Next, we excluded samples with missing values in BMI or RACE. If RACE or BMI were not already in binary format, we converted them using a standardized approach: RACE was coded as 1 for white race and 0 for all other races, while BMI was coded as 0 for normal values (18.5--24.9) and 1 for abnormal values (outside this range).

In addition, for two of the studies containing multi-center datasets, we divided the data by region to create separate datasets, thereby increasing the number of sample groups from 6 to 12. Finally, we calculated the approximate value of \( p_{11} \) based on all the individual data from these 12 datasets and transformed each dataset into study-level data $(n_i, x_i, y_i)$ where $i=1,...,12$. Relevant information about these 12 datasets is provided in Table \ref{real_data}.

\end{document}